\documentclass[12pt,a4paper,final]{article}
\usepackage[latin1]{inputenc}
\usepackage{a4wide}
\usepackage{amssymb}
\usepackage[notref,notcite]{showkeys}
\usepackage{ifpdf}
\ifpdf
\usepackage[pdftex]{graphicx}
\usepackage[pdftex,unicode,implicit]{hyperref}
\hypersetup{%
  pdftitle    = {All the supersymmetric solutions of N=1,d=5 
ungauged supergravity},
  pdfkeywords = {Supersymmetry, real special geometry, BPS solutions},
  pdfauthor   = {J. Bellorín, P. Meessen, T. Ortín},
  pdfcreator  = {pdf\LaTeXe\ with package \flqq hyperref\frqq},
  pdfproducer = {pdf\LaTeXe\ with package \flqq hyperref\frqq},
  pdfpagemode = None,  
  pdffitwindow= true,  
  unicode     = true,
  plainpages  = true,
  colorlinks  = true,  
  citecolor   = blue,
  urlcolor    = red,
  linkcolor   = black
}
\newcommand{\hepth}[1]{arXiv:{\tt \href{http://www.arXiv.org/abs/hep-th/#1}{hep-th/#1}}}

\else
  \usepackage[dvips]{graphicx}
  \usepackage[unicode,implicit]{hyperref}
  \newcommand{\hepth}[1]{arXiv:{\tt hep-th/#1}}

\fi
\makeatletter
\@addtoreset{equation}{section}
\makeatother

\begin{document}
\begin{flushright}
\small
IFT-UAM/CSIC-06-38\\
{\bf hep-th/0610196}\\
October $17^{\rm th}$, $2006$
\normalsize
\end{flushright}
\begin{center}
\vspace{2cm}
{\LARGE {\bf All the supersymmetric solutions of}}\\[.5cm]
{\LARGE {\bf $N=1,d=5$ ungauged supergravity}} 
\vspace{2cm}

{\sl\large Jorge Bellor\'{\i}n}
\footnote{E-mail: {\tt Jorge.Bellorin@uam.es}},
{\sl\large Patrick Meessen}
\footnote{E-mail: {\tt Partick.Meessen@cern.ch}},
{\sl\large and Tom{\'a}s Ort\'{\i}n}
\footnote{E-mail: {\tt Tomas.Ortin@cern.ch}}

\vspace{1cm}

{\it Instituto de F\'{\i}sica Te\'orica UAM/CSIC\\
Facultad de Ciencias C-XVI,  C.U.~Cantoblanco,  E-28049-Madrid, Spain}\\

\vspace{2cm}


{\bf Abstract}

\end{center}

\begin{quotation}\small
  We classify the supersymmetric solutions of ungauged $N=1$ $d=5$ SUGRA
  coupled to vector multiplets and hypermultiplets.  All the solutions can be
  seen as deformations of solutions with frozen hyperscalars. We show
  explicitly how the 5-dimensional Reissner-Nordstr\"om black hole is deformed
  when hyperscalars are living on $SO(4,1)/SO(4)$ are turned on, reducing its
  supersymmetry from $1/2$ to $1/8$. We also describe in the timelike and null
  cases the solutions that have one extra isometry and can be reduced to
  $N=2,d=4$ solutions. Our formulae allows the uplifting of certain $N=2,d=4$
  black holes to $N=1,d=5$ black holes on KK monopoles or to pp-waves
  propagating along black strings.
\end{quotation}

\newpage

\pagestyle{plain}


\tableofcontents

\newpage

\section{Introduction}

With Ref.~\cite{Gauntlett:2002nw}, Gauntlett {\em et al.\/} revolutionized the
art of finding supersymmetric solutions, by extending the methods pioneered by
Tod \cite{Tod:1983pm} and applying them to classify the supersymmetric
solutions of minimal $N=1$ $d=5$ supergravity. Since then, there has been a
renewed, vigorous and systematic effort in the literature to classify, or at
least characterize, generic supersymmetric solutions of supergravity theories.
In the framework of $N=1,d=5$ SUGRA the results of
Ref.~\cite{Gauntlett:2002nw} were extended to the gauged case in
Ref.~\cite{Gauntlett:2003fk}, to include the coupling to an arbitrary number
of vector multiplets in Ref.~\cite{Gauntlett:2004qy} and their Abelian
gaugings were further considered in
Refs.~\cite{Gutowski:2004yv,Gutowski:2005id}\footnote{Previous work on these
  theories can be found in Refs.~\cite{Chamseddine:1998yv,Sabra:1997yd}.}. In
the framework of $N=2,d=4$ SUGRA the new methods allowed the extension of
Tod's results to pure gauged $N=2,d=4$ SUGRA \cite{Caldarelli:2003pb} and to
ungauged $N=2,d=4$ SUGRA coupled to an arbitrary number of vector multiplets
\cite{Meessen:2006tu} and hypermultiplets \cite{Hubscher:2006mr}. The minimal
$d=6$ SUGRA was dealt with in Refs.~\cite{Gutowski:2003rg,Chamseddine:2003yy},
some gaugings were considered in Ref.~\cite{Cariglia:2004kk} and the coupling
to hypermultiplets has been fully solved in Ref.~\cite{Jong:2006za}.  Further
works in other (higher) dimensions and number of supercharges or based on the
alternative methods of spinorial geometry are
Refs.~\cite{Gauntlett:2002fz,Bellorin:2005zc}.

In this paper we will extend further the results obtained in ungauged
$N=1,d=5$ SUGRA to include, on top of vector multiplets, hypermultiplets. This
problem was considered before by Cacciatori, Celi and Zanon in
Refs.~\cite{Cacciatori:2002qx,Celi:2003qk,Cacciatori:2004qm}, making progress
towards a full solution of the problem which we present here. 

Similar works in 4- and 6-dimensional SUGRAs with 8 supercharges ($N=2,d=4$
and $N=(1,0),d=6$) coupled to vector multiplets and hypermultiplets have been
recently published \cite{Hubscher:2006mr,Jong:2006za}.  As the observant
reader will see, there is a staggering similarity between the results found in
those works and the ones presented here.  The reason for this is simply
because the hypermultiplets have a very characteristic, and minimal, way of
coupling to the rest of the fields, a coupling that is roughly the same in the
3 theories with 8 supercharges, wherefore the resulting structures should be
comparable.

Let us describe our results qualitatively: all the supersymmetric solutions
can be seen as deformations of supersymmetric solutions with the same electric
and magnetic charges but frozen hyperscalars (which is effectively the same
as having only vector
multiplets), which were classified in Ref.~\cite{Gauntlett:2003fk}. The effect
of defrosting the hyperscalars is an electric and magnetic charge preserving 
deformation of those solutions;
the deformations consist in a
deformation of the base space in the timelike case and of the wavefront space
in the null case. To be more precise, in the timelike case, the metrics of all
the supersymmetric solutions have the general conformastationary form

\begin{equation}
ds^{2} = f^{2}\left(dt+\omega\right)^{2}
-f^{-1}h_{\underline{m}\underline{n}} dx^{m}dx^{n}\, .
\end{equation}

\noindent
$h_{\underline{m}\underline{n}}$ is the time-independent base space metric
and when dealing with frozen hypermultiplets, it has to be hyper-K\"ahler. The metric, with
$f=1$ and $\omega=0$ and vanishing matter fields is a supersymmetric solution
by itself and can be seen as a background which is excited when electric and
magnetic charges are turned on.  The functions $f$ and $\omega$ are
essentially determined by the electric and magnetic charges and satisfy
covariant differential equations in the base space.

When the hyperscalars are turned on $h_{\underline{m}\underline{n}}$ is no
longer a hyper-K\"ahler manifold: the form of this metric is dictated by two
requirements

\begin{enumerate}
\item The hyperscalars $q^{X}(x)$ are quaternionic maps\footnote{
          Please see the discussion after Eq. (\ref{triholomorphic}) for more information about 
          the notion of quaternionic maps.}
      from the base space to the quaternionic-K\"ahler target manifold.
\item The anti-selfdual part of the spin connection of the base manifold has
  to be equal (up to gauge transformations) to the pullback of the $\mathfrak{su}(2)$
  connection characterizing the quaternionic-K\"ahler target manifold.
\end{enumerate}

\noindent
These two conditions are interwoven but, as we will show in an explicit
example, can be solved simultaneously.  

Now, the metric, with $f=1$ and $\omega=0$, vanishing vector multiplets but
unfrozen hyperscalars is a supersymmetric solution by itself and can be seen
as a background which is excited when electric and magnetic charges are turned
on. The functions $f$ and $\omega$ satisfy the same covariant differential
equations as before but in the new base space metric. 

These solutions generically preserve only $1/8$ of the available $8$
supersymmetries. 

In the null case, the metric is generically of the form

\begin{equation}
ds^{2}= 2fdu(dv+Hdu+\omega)
-f^{-2}\gamma_{\underline{r}\underline{s}}dx^{r}dx^{s}\, ,  
\end{equation}

\noindent
where $r,s=1,2,3$ and all functions are $v$-independent. The functions $f$ and
$H$ and the 1-form $\omega$ depend on the electric and magnetic charges and
satisfy differential equations in the background of the 3-dimensional
wavefront metric $\gamma_{\underline{r}\underline{s}}$. When the hyperscalars
are frozen, this metric is flat;  when they are turned on, the 3-dimensional
metric is determined by exactly the same two conditions that the base space of
supersymmetric solutions of $N=2,d=4$ SUGRA coupled to hypermultiplets
satisfy, namely

\begin{enumerate}
\item The hyperscalars must satisfy

\begin{equation}
\partial_{r}q^{X}\ f_{X}{}^{iA}\ \sigma^{r}{}_{i}{}^{j}\ =\ 0\, .  
\end{equation}

\item The spin connection of the 3-dimensional metric must be equal (up to
  gauge transformations) to the pull-back of the the $\mathfrak{su}(2)$ connection that
  characterizes the quaternionic-K\"ahler target manifold.
\end{enumerate}

This suggests a relation with the 4-dimensional solutions. We thus consider
the particular case in which the metric has an additional isometry and is, in
particular, $u$-independent. It is not difficult to see that in general the solutions of
the null case describe pp-waves propagating along a string.
Solutions which are $u$-independent can be compactified along the direction in
which the wave propagates, \textit{i.e.}~along the string and give solutions
belonging to the 4-dimensional timelike class, \textit{i.e.}~black hole-type
solutions. 

This set of 5-dimensional solutions and their reductions are presented here
for the first time and allow an uplifting of 4-dimensional black-hole-type
solutions (with or without hypermultiplets) to $d=5$ dimensions different from
the one considered in
Refs.~\cite{Bena:2004tk,Bena:2005ay,Gaiotto:2005gf,Elvang:2005sa,Gaiotto:2005xt,Bena:2005ni,Behrndt:2005he}.
There, 4-dimensional black holes were uplifted to 4-dimensional black holes in
a KK monopole background.  Here we are dealing with the electric-magnetic dual
uplift since the simplest $5$-dimensional pp-wave and the Sorkin-Gross-Perry
KK monopole \cite{Sorkin:1983ns} are related by dimensional reduction to $d=4$
dimensions and $4$-dimensional electric-magnetic duality, the 4-dimensional
solution being the so-called ``KK black hole'', which in this simple case is
singular.  This relation is known in the general case under the name of
``$r$-map'', whence the $r$-map will relate these new string-pp-wave
upliftings\footnote{A particular case of this kind of uplifting was also
  observed in Ref.~\cite{Cvetic:1998xh}, although the 5-dimensional solutions
  were interpreted as rotating strings.} to the known black hole-KK monopole
upliftings. 

This uplift may be more convenient to understand the black hole solutions from
a higher-dimensional point of view since they are direct realizations of the
D1-D5-W model. It may shed light on Mathur's conjecture
\cite{Lunin:2001jy,Mathur:2005zp} on the realization of D1-D5-W microstates as
supergravity solutions \cite{Lunin:2002qf}.

For the sake of completeness we have also worked out the timelike case with
one additional isometry as, with frozen hyperscalars, all of the
interesting solutions (supersymmetric rotating black holes and black rings
\cite{Elvang:2004rt}) seem to belong to this class
\cite{Gauntlett:2002nw,Gauntlett:2004wh,Gauntlett:2004qy}. The base space
manifold is now a generalization of the Gibbons-Hawking instanton metric
\cite{Gibbons:1979zt}. The Gibbons-Hawking instanton metric is the most
general 4-dimensional hyper-K\"ahler metric with one isometry and can be used
as a base space metric $h_{\underline{m}\underline{n}}$ in absence of
hyperscalars. It has the form

\begin{equation}
ds^{2}_{(4)} =  H^{-1} (dz +\chi)^{2}
+H \delta_{\underline{r}\underline{s}}dx^{r}dx^{s}\, ,\,\,\,\, r,s=1,2,3\, ,
\end{equation}

\noindent
where $H$ is a function harmonic on 3-dimensional Euclidean space. 

In presence of unfrozen hyperscalars the metric to be considered is

\begin{equation}
ds^{2}_{(4)} =  H^{-1} (dz +\chi)^{2}
+H \gamma_{\underline{r}\underline{s}}dx^{r}dx^{s}\, ,\,\,\,\, r,s=1,2,3\, ,
\end{equation}

\noindent where the spin connection of the 3-dimensional metric
$\gamma_{\underline{r}\underline{s}}$ has to be equal (up to gauge
transformations) to the pullback of the $\mathfrak{su}(2)$ connection of the hyperscalar
manifold.


This paper is organized as follows: in Section~\ref{sec-n1d5m} we describe
ungauged $N=1,d=5$ supergravity coupled to vector multiplets and
hypermultiplets. In Section~\ref{sec-ksis} we derive the integrability
conditions (KSIs) of the Killing spinor equations (KSEs), that relate the
equations of motion of the fields for supersymmetric configurations, which
will allow us to minimize the number of independent equations that need to be
solved. In Section~\ref{sec-susy} we proceed to find the supersymmetric
configurations and solutions, both in the timelike,
(Section~\ref{sec-timelike},) and in the null (Section~\ref{sec-null})
classes.  An explicit example of the timelike class with unfrozen hyperscalars
is given in Section~\ref{sec:TLEx} and the general subclasses of solutions
that generically have one additional isometry are given in
Sections~\ref{sec-timelikeisometry} (timelike case) and
Section~\ref{sec-nullisometry} (null case).  Section~\ref{sec-conclusions}
contains our conclusions and final thoughts.  Appendix~\ref{sec-d5conventions}
contains our conventions on gamma matrices, spinors, spinor bilinears and real
special geometry.  Appendix~\ref{sec-quaternionic} contains a brief
introduction to quaternionic-K\"ahler manifolds. Finally,
Appendices~\ref{app-metric} and \ref{app-nullmetric} contain the necessary
geometric data for the 5-dimensional metrics that appear in this paper.



\section{Matter-coupled, ungauged $N=1,d=5$ supergravity}
\label{sec-n1d5m}

In this section we describe briefly the supergravity theories we will be
working with: $N=1,d=5$ (minimal) ungauged supergravity coupled to $n_{v}$
vector multiplets and $n_{h}$ hypermultiplets\footnote{We follow essentially
  the notation and conventions of Ref.~\cite{Bergshoeff:2004kh} with some
  minor changes to adapt them to those in
  Refs.~\cite{Lozano-Tellechea:2002pn,Meessen:2004mh}. The changes are
  explained in Appendix~\ref{sec-d5conventions}. The original references on
  matter-coupled $N=1,d=5$ SUGRA are \cite{Gunaydin:1983bi} and
  \cite{Gunaydin:1984ak}. The origin of these theories from compactifications
  of 11-dimensional supergravity on Calabi-Yau 3-folds was studied in
  Ref.~\cite{Cadavid:1995bk}.}.

The supergravity multiplet consists of the graviton $e^{a}{}_{\mu}$, the
graviphoton $A_{\mu}$ and the gravitino $\psi_{\mu}^{i}$. The gravitino and
the rest of spinors in the theory are pairs of symplectic-Majorana spinors
$i=1,2$ as explained in Appendix~\ref{app-spinors}.  

Each of the $n_{v}$ vector multiplets, labeled by $x=1,\cdots ,n_{v}$ consists
of one real vector field $A_{\mu}^{x}$, a real scalar $\phi^{x}$ and a gaugino
$\lambda^{xi}$. The scalars $\phi^{x}$, parametrize a Riemannian manifold
which we call ''the scalar manifold". The full theory is formally invariant
under an $SO(n_{v}+1)$ symmetry that mixes the matter vectors $A^{x}{}_{\mu}$
with the supergravity vector $A_{\mu}\equiv A^{0}{}_{\mu}$ and so it is
convenient to treat all the vector fields on the same footing denoting them by
$A^{I}{}_{\mu}$ $I=0,\cdots, n_{v}$. The symmetry that rotates the vectors acts
on the scalars as well and, to make it manifest one defines $n_{v}+1$
functions of the physical scalars $h^{I}(\phi)$. These functions satisfy the
constraint

\begin{equation}
C_{IJK}h^{I}h^{J}h^{K}=1\, ,  
\end{equation}

\noindent
where $C_{IJK}$ is a fully symmetric real constant tensor which characterizes
completely the couplings in the vectorial sector. In particular it determines
the metric of the scalar manifold $g_{xy}(\phi)$ on the target of $\phi^{x}$,
the couplings between scalars and vector fields $a_{IJ}(\phi)$ and the
coupling constants of the vector field Chern-Simons terms. The relations
between these fields are given in the Appendix~\ref{sec-realspecialgeometry}.

Each of the $n_{h}$ hypermultiplets consists of four real scalar-fields
(\textit{hyperscalars}) $q^{X}$, $X=1,\cdots,4n_{h}$ and two spinor fields
(\textit{hyperinos}) $\zeta^{A}$, $A=1,\ldots,2n_{h}$. The index $i$
associated to the symplectic-Majorana condition is embedded into the index
$A$.  The hyperscalars $q^{X}$ parametrize a quaternionic-K\"ahler manifold,
described in Appendix~\ref{sec-quaternionic}, that we will refer to
as the {\em hypervariety}.  In particular we observe that the connection of
quaternionic-K\"ahler manifolds can be decomposed in an
$\mathfrak{sp}(1)\simeq\mathfrak{su}(2)$ and an $\mathfrak{sp}(n_{h})$
component whose pullback to spacetime will act on objects with index $i$ and
$A$, respectively.

The bosonic part of the action is

\begin{equation}
  \begin{array}{rcl}
S & = &  {\displaystyle\int} d^{5}x\sqrt{g}\
\biggl\{
R
+{\textstyle\frac{1}{2}}g_{xy}\partial_{\mu}\phi^{x}\partial^{\mu}\phi^{y}
+{\textstyle\frac{1}{2}}g_{XY}\partial_{\mu}q^{X}\partial^{\mu}q^{Y} 
\\
& & \\
& & 
\hspace{2cm}
-{\textstyle\frac{1}{4}} a_{IJ} F^{I\, \mu\nu}F^{J}{}_{\mu\nu}
+{\textstyle\frac{1}{12\sqrt{3}}}C_{IJK}
{\displaystyle\frac{\varepsilon^{\mu\nu\rho\sigma\alpha}}{\sqrt{g}}}
F^{I}{}_{\mu\nu}F^{J}{}_{\rho\sigma}A^{K}{}_{\alpha}
\biggr\}\, .
\end{array}
\end{equation}

Observe that the hyperscalars do not couple to any of the fields in the vector
multiplets and couple to the supergravity multiplet only through the metric.
This is similar to what happens in $N=2,d=4$ theories and will have similar
consequences.

We use the following notation for the equations of motion

\begin{equation}
\mathcal{E}_{a}{}^{\mu}\equiv 
-\frac{1}{2\sqrt{g}}\frac{\delta S}{\delta e^{a}{}_{\mu}}\, ,
\,\,\,\,\,
\mathcal{E}_{x} \equiv -\frac{1}{\sqrt{g}}
\frac{\delta S}{\delta \phi^{x}}\, ,
\,\,\,\,\,
\mathcal{E}_{X} \equiv -\frac{1}{\sqrt{g}}
\frac{\delta S}{\delta q^{X}}\, ,
\,\,\,\,\,
\mathcal{E}_{I}{}^{\mu}\equiv 
\frac{1}{\sqrt{g}}\frac{\delta S}{\delta A^{I}{}_{\mu}}\, ,
\end{equation}

\noindent
which are given by

\begin{eqnarray}
\mathcal{E}_{\mu\nu} 
& = & 
G_{\mu\nu}
-{\textstyle\frac{1}{2}}a_{IJ}\left(F^I{}_{\mu}{}^{\rho} F^{J}{}_{\nu\rho}
-{\textstyle\frac{1}{4}}g_{\mu\nu}F^{I\, \rho\sigma}F^{J}{}_{\rho\sigma}
\right)      
+{\textstyle\frac{1}{2}}g_{xy}\left(\partial_{\mu}\phi^{x}\partial_{\nu}\phi^{y}
-{\textstyle\frac{1}{2}}g_{\mu\nu}
\partial_\rho\phi^{x} \partial^{\rho}\phi^{y}\right)
\nonumber\\
& & \nonumber  \\
& & 
+{\textstyle\frac{1}{2}}g_{XY}\left(\partial_{\mu}q^{X}\partial_{\nu}q^{Y}
-{\textstyle\frac{1}{2}}g_{\mu\nu} \partial_{\rho}q^{X}\partial^{\rho}q^{Y}
\right)\, ,\label{eq:Emn} \\
& & \nonumber \\
g^{xy}\mathcal{E}_{y} 
& = & 
\mathfrak{D}_{\mu}\partial^{\mu}\phi^{x} 
+{\textstyle\frac{1}{4}} g^{xy}\partial_{y} 
a_{IJ} F^{I\, \rho\sigma}F^{J}{}_{\rho\sigma}\, , \label{eq:Ei}\\
& & \nonumber \\
g^{XY}\mathcal{E}_{Y} 
& = & 
\mathfrak{D}_{\mu}\partial^{\mu}q^{X}\, , \label{eq:EX}\\
& & \nonumber \\
\mathcal{E}_{I}{}^{\mu} 
& = & 
\nabla_{\nu}(a_{IJ}F^{J\, \nu\mu})
+{\textstyle\frac{1}{4\sqrt{3}}} C_{IJK} 
\frac{\varepsilon^{\mu\nu\rho\sigma\alpha}}{\sqrt{g}}
F^{J}{}_{\nu\rho}F^{J}{}_{\sigma\alpha}\, .
\label{eq:ERm}
\end{eqnarray}

To these definitions we add the following notation for the Bianchi identities
of the vector fields:

\begin{equation}
\mathcal{B}^{I}{}_{\mu\nu\rho} \equiv 3\nabla_{[\mu}F^{I}{}_{\nu\rho]}\, .  
\end{equation}

In these equations $\mathfrak{D}_{\mu}$ is the covariant derivative in the
spacetime and in the corresponding scalar manifold. Then, Eq.~(\ref{eq:EX})
states that $q$ is a harmonic map from spacetime to the hypervariety.

The supersymmetry transformation rules for the fermionic fields, evaluated on
vanishing fermions, are

\begin{eqnarray}
\delta_{\epsilon}\psi^{i}_{\mu} 
& = & 
\mathsf{D}_{\mu}\epsilon^{i}
-{\textstyle\frac{1}{8\sqrt{3}}}h_{I}F^{I\,\alpha\beta}
\left(\gamma_{\mu\alpha\beta}-4g_{\mu\alpha}\gamma_\beta\right)
\epsilon^{i}\, , \\
& & \nonumber \\ 
\delta_{\epsilon}\lambda^{ix} 
& = &  
{\textstyle\frac{1}{2}}\left(\not\!\partial\phi^{x} 
-{\textstyle\frac{1}{2}}h^{x}_{I}\not\!F^{I}\right)\epsilon^{i}\, ,\\
& & \nonumber \\
\delta_{\epsilon}\zeta^{A} 
& = & 
{\textstyle\frac{1}{2}}f_{X}{}^{iA}\not\!\partial q^{X}\epsilon_{i}\, ,
\end{eqnarray}

\noindent
where $\mathsf{D}_{\mu}$ is the Lorentz- and $SU(2)$-covariant derivative

\begin{equation}
\mathsf{D}_{\mu}\epsilon^{i} \equiv \nabla_{\mu}\epsilon^{i} 
 +\epsilon^{j}\mathsf{A}_{j}{}^{i}{}_{\mu}\, ,  
\end{equation}

\noindent
and the $\mathfrak{su}(2)$ connection is the pullback of the
$\mathfrak{su}(2)$ connection of the hypervariety:

\begin{equation}
\mathsf{A}^{r}{}_{\mu} \;\equiv\;  \partial_{\mu}q^{X}\ \omega_{X}{}^{r}\, , 
\hspace{1cm}
 \mathsf{A}_{j}{}^{i} \; =\; i\mathsf{A}^{r}\ \sigma^{r}{}_{j}{}^{i}\, .
\end{equation}

Observe that the hyperscalars only appear in the gravitino's and gauginos'
supersymmetry transformation rules precisely through the $\mathfrak{su}(2)$
connection.

Finally, the supersymmetry transformation rules of the bosonic fields are

\begin{eqnarray}
\label{eq:susytranse}
\delta_{\epsilon} e^{a}{}_{\mu} 
& = & 
{\textstyle\frac{i}{2}} \bar{\epsilon}_{i}\gamma^a\psi^{i}_{\mu}\, ,
\\
& & \nonumber \\ 
\label{eq:susytransA}
\delta_{\epsilon} A^{I}{}_{\mu} 
& = & 
-{\textstyle\frac{i\sqrt{3}}{2}}h^{I}\bar{\epsilon}_{i}\psi^{i}_{\mu}
+{\textstyle\frac{i}{2}}h^{I}_{x}\bar{\epsilon}_{i}\gamma_{\mu}\lambda^{x\,i}
\, ,
\\
& & \nonumber \\
\label{eq:susytransZ}
\delta_{\epsilon} \phi^{x} 
& = & 
{\textstyle\frac{i}{2}}\bar{\epsilon}_{i}\lambda^{x\,i}\, ,
\\
& & \nonumber \\
\label{eq:susytransq}
\delta_{\epsilon} q^{X} 
& = & 
-if_{iA}{}^{X}\bar{\epsilon}^{i}\zeta^{A}\, .
\end{eqnarray}


\section{KSIs and integrability conditions}
\label{sec-ksis}

The bosons' supersymmetry transformation rules lead to the following KSIs
\cite{Kallosh:1993wx,Bellorin:2005hy} associated to the
gravitino, gauginos and hyperinos {\em resp.\/}:

\begin{eqnarray}
\label{eq:preksi1}
 \left(
\mathcal{E}_{\mu}{}^\nu\gamma_{\nu} 
+{\textstyle\frac{\sqrt{3}}{2}}h^{I}\mathcal{E}_{I\, \mu}
\right)\epsilon^{i} 
& = & 
0\, ,\\
& & \nonumber\\
\label{eq:preksi2}
\left(\mathcal{E}_{x} -h^{I}_{x} \not\!\mathcal{E}_{I}\right)\epsilon^{i}
& = & 
0\, , \\
& &  \nonumber\\
\label{eq:ksi2b}
f_{iA}{}^{X}\mathcal{E}_{X}\epsilon^{i} & = & 0\,.
\end{eqnarray}

\noindent
It is an implicit assumption, used to derive the KSIs, that the Bianchi
identities are satisfied. This affects, in particular, the first two KSIs,
where the vector field equations appears. It is, therefore, useful to derive
them from the integrability conditions of the KSEs, even if the derivation
requires much more work, because in this case, contrary to what happens in
$N=2,d=4$ theories \cite{Meessen:2006tu}, there is no electric-magnetic
symmetry indicating in what combination the Bianchi identities should
accompany the Maxwell equations.
 
The integrability condition of the KSE associated to the gravitino
supersymmetry transformation gives

\begin{equation}
  \begin{array}{rcl}
4\gamma^{\nu}\mathsf{D}_{[\mu}\delta_{\epsilon}\psi^{i}_{\nu]}
& =  &
\left\{
\left(\mathcal{E}_{\mu}{}^{\sigma}-{\textstyle\frac{1}{3}}
g_{\mu}{}^{\sigma}\,\mathcal{E}_{\rho}{}^{\rho}\right)\gamma_{\sigma}
\right. 
\\
& & \\
& & 
\left.
+{\textstyle\frac{1}{4\sqrt{3}}}h^{I}
\left[\gamma_{\mu}\left(\not\!\mathcal{E}_{I}
+{\textstyle\frac{1}{6}a_{IJ}}\not\!\mathcal{B}^{J}\right)
+3\left(\not\!\mathcal{E}_{I}
+{\textstyle\frac{1}{6}a_{IJ}}\not\!\mathcal{B}^{J}\right)
\gamma_{\mu}\right]
\right\}
\epsilon^{i}=0\, .
\end{array}
\end{equation}

To obtain this equation we need to use Eqs.~(\ref{eq:una})-(\ref{eq:otra}),
with $\nu=-1$ as to ensure the correct normalization of the hyperscalars'
energy-momentum tensor. It is a well-known result that manifolds with the
opposite sign of $\nu$ cannot be coupled to supergravity and here we are just
recovering this result.

Acting with $\gamma^{\mu}$ from the left, we get 

\begin{equation}
\left[\mathcal{E}_{\rho}{}^{\rho} + {\textstyle\frac{\sqrt{3}}{2}} 
h^{I}(\not\!\mathcal{E}_{I}
-{\textstyle\frac{1}{3}}a_{IJ}\not\!\mathcal{B}^{J})\right]
\epsilon^{i}=0\, ,
\end{equation}

\noindent
which can be used to eliminate $\mathcal{E}_{\rho}{}^{\rho}$ from the
integrability equation:

\begin{equation}
\label{eq:ksi1}
\left[\left( \mathcal{E}_{\mu}{}^{\sigma}
+{\textstyle\frac{\sqrt{3}}{2}}
h_{I}{}^{\star}\mathcal{B}^{I}{}_{\mu}{}^{\sigma}\right)
\gamma_{\sigma} +{\textstyle\frac{\sqrt{3}}{2}}h^{I}\mathcal{E}_{I\, \mu}
\right]\epsilon^{i} = 0\, .
\end{equation}

On the other hand, from the gauginos' supersymmetry transformation rule
we get

\begin{equation}
\label{eq:ksi2}
        2\not\!\!\mathfrak{D}\delta_{\epsilon}\lambda^{ix} =  
        \left[\mathcal{E}_{x} -h^{I}_{x} \left(\not\!\mathcal{E}_{I}
        +{\textstyle\frac{1}{6}}a_{IJ}\not\!\mathcal{B}^{J}\right) \right]
        \epsilon^{i}= 0\, .
\end{equation}

Eqs.~(\ref{eq:ksi1}) and (\ref{eq:ksi2}) are the modifications to the two KSIs
Eq. ~(\ref{eq:preksi1}) and Eq.~(\ref{eq:preksi2}) that we were seeking for.

Let us now obtain tensorial equations form the spinorial KSIs: acting with
$i\bar{\epsilon}_{i}\gamma_{\rho}$ from the left on Eq.~(\ref{eq:ksi1}) and
taking into account the properties of the spinor bilinears discussed in
Appendix~\ref{sec-bilinears}, we get

\begin{equation}
\label{eq:ksi3}
f\left( \mathcal{E}_{\mu\rho}
+{\textstyle\frac{\sqrt{3}}{2}}
h_{I}{}^{\star}\mathcal{B}^{I}{}_{\mu\rho}\right)
+{\textstyle\frac{\sqrt{3}}{2}}h^{I}\mathcal{E}_{I\, \mu}V_{\rho}
 = 0\, ,
\end{equation}

\noindent
whose symmetric and antisymmetric parts give independent equations.

Doing the same on Eqs.~(\ref{eq:ksi2}) and~(\ref{eq:ksi2b}), we get 

\begin{eqnarray}
\label{eq:ksi6}
\mathcal{E}_{x}V^{\rho} - fh^{I}_{x} \mathcal{E}_{I}{}^{\rho} 
& = & 
0\, , \\
& & \nonumber \\
\label{eq:ksi6b}
\mathcal{E}_{X}V^{\rho} & = & 0\, .
\end{eqnarray}

\noindent
Finally, acting with $i\bar{\epsilon}_{i}$ on Eqs.~(\ref{eq:ksi1}),
(\ref{eq:ksi2}) and~(\ref{eq:ksi2b}) from the left we get respectively

\begin{eqnarray}
\label{eq:ksi7}
\left( \mathcal{E}_{\mu\rho}
+{\textstyle\frac{\sqrt{3}}{2}}
h_{I}{}^{\star}\mathcal{B}^{I}{}_{\mu\rho}\right)V^{\rho}
+{\textstyle\frac{\sqrt{3}}{2}}fh^{I}\mathcal{E}_{I\, \mu}
 & =  & 0\, ,\\
& & \nonumber \\
\label{eq:ksi72}
f\mathcal{E}_{x}-h^{I}_{x} \mathcal{E}_{I\, \rho}V^{\rho} & = & 0\, , \\
& & \nonumber \\
\mathcal{E}_{X} f & = & 0\,.
\end{eqnarray}

\noindent
which can be obtained from Eqs.~(\ref{eq:ksi3})-(\ref{eq:ksi6b}) only in the
timelike $f\neq 0$ case.

Summarizing, in the timelike case, defining the unimodular timelike vector
$v^{\mu}\equiv V^{\mu}/f$, we have

\begin{eqnarray}
\label{eq:ksi8}
\mathcal{E}^{\mu\nu} 
& = & 
-{\textstyle\frac{\sqrt{3}}{2}}h^{I} \mathcal{E}_{I}{}^{(\mu}v^{\nu)}\, ,\\
& & \nonumber \\
\label{eq:ksi9}
h_{I}{}^{\star}\mathcal{B}^{I\, \mu\nu} 
& = &
-h^{I}\mathcal{E}_{I}{}^{[\mu}v^{\nu]}\, , \\
& & \nonumber \\
\label{eq:ksi10}
\mathcal{E}_{x} &=& h^{I}_{x} \mathcal{E}_{I}{}_{\rho} v^{\rho} \,, \\
& & \nonumber \\
\label{eq:ksi10b}
\mathcal{E}_{X} &=& 0\, ,
\end{eqnarray}

\noindent
which imply that all the supersymmetric configurations automatically solve the
equation of motion of the hyperscalars and that, if the Maxwell equations are
satisfied, then the Einstein and scalar equations and the projections
$h_{I}\mathcal{B}^{I}$ of the Bianchi identities are also satisfied.
Therefore, in the timelike case, the necessary and sufficient condition for a
supersymmetric configuration to also be a solution of the theory, is that it
must solve the Maxwell equations and the Bianchi identities. Observe that,
contrary to the 4-dimensional cases in which only one component of the Maxwell
equations and Bianchi identities (the time component) need to be checked
because the rest are automatically satisfied, in this 5-dimensional case we
need to check all the components of the Maxwell equations and of the Bianchi
identities.



In the null ($f=0$) case, we get, renaming $V^{\mu}$ as $l^{\mu}$ 

\begin{eqnarray}
\label{eq:ksi11}
\mathcal{E}_{\mu\rho}l^{\rho}
& = & 
-{\textstyle\frac{\sqrt{3}}{2}}
h_{I}{}^{\star}\mathcal{B}^{I}{}_{\mu\rho}l^{\rho}\, ,\\
& & \nonumber \\
\label{eq:ksi12}
h^{I}\mathcal{E}_{I\, \mu} & = & 0\, ,\\
& & \nonumber \\
\label{eq:ksi13}
h^{I}_{x}\mathcal{E}_{I\, \rho}l^{\rho} & = & 0\, ,\\
& & \nonumber \\
\label{eq:ksi14}
\mathcal{E}_{x} & = & 0\, ,     \\
& & \nonumber \\
\label{eq:ksi15}
\mathcal{E}_{X} &=& 0\,,
\end{eqnarray}

\noindent
which imply that the scalar and hyperscalars equations are automatically
satisfied and so are certain projections of the Maxwell and Einstein
equations.


\section{Supersymmetric configurations and solutions}
\label{sec-susy}

In this section we will follow the procedure of Ref.~\cite{Gauntlett:2002nw}
to obtain supersymmetric configurations of supergravity, which consists in
deriving equations for all the bilinears that can be constructed from the
Killing spinors. These equations contain the lion's part of the information
contained in the KSEs and can be used to constrain the form of the bosonic
fields. These constraints are necessary conditions for the configurations to
be supersymmetric and subsequently one has to prove that they are also
sufficient (or find the missing conditions, as will happen in the null case).
Finally one has to impose the equations of motion on the supersymmetric
configurations in order to have classical supersymmetric solutions. The KSIs,
derived in the previous section, simplify this task since only a small number
of equations of motion are independent for supersymmetric configurations.

As we remarked in section~\ref{sec-n1d5m}, the hyperscalars appear only
implicitly in the gravitino and gauginos supersymmetry transformations through
the pullback of the $\mathfrak{su}(2)$ connection. The equations we are going
to obtain for the fields in the supergravity and vector multiplets are,
therefore, formally identical to the case without hypermultiplets considered
in Ref.~\cite{Gutowski:2004yv}, but containing implicitly the
$\mathfrak{su}(2)$ connection and its consequences.  This is similar to what
happens in the coupling of $N=2,d=4$ theories to hypermultiplets considered
only recently in Ref.~\cite{Hubscher:2006mr}

Our goal is to find all the field configurations for which the KSEs 

\begin{eqnarray}
\label{gravitinokse}
\left\{
\mathsf{D}_{\mu}-{\textstyle\frac{1}{8\sqrt{3}}}h_{I}F^{I\,\alpha\beta}
\left(\gamma_{\mu\alpha\beta}-4g_{\mu\alpha}\gamma_\beta\right)
\right\}
\epsilon^{i} 
& = & 0\, , \\
& & \nonumber \\
\label{gauginokse} 
\left(\not\!\partial\phi^{x} 
-{\textstyle\frac{1}{2}}h^{x}_{I}\not\!F^{I}\right)\epsilon^{i} 
& = & 0\, ,\\
& & \nonumber \\
\label{hyperinokse}
f_{X}{}^{iA}\!\not\!\partial q^{X}\epsilon_{i} 
& = & 0\, ,
\end{eqnarray}

\noindent
admit at least one solution $\epsilon^{i}$. We are going to assume its
existence and we are going to derive necessary conditions for this to happen.
These conditions will arise as consistency conditions of the equations
satisfied by the tensors that can be constructed as bilinears of the Killing
spinor whose existence was assumed from the onset. 

As explained in Appendix~(\ref{sec-bilinears}), the tensor-bilinears that can
be constructed from a symplectic-Majorana spinor are a scalar $f$, a vector
$V$ and three 2-forms $\Phi^{r}$.  $f$ and $V$ are $SU(2)$-singlets whereas
the $\Phi$s form an $SU(2)$-triplet.

The fact that the Killing spinor satisfies Eq.~(\ref{gravitinokse}) leads to
the following differential equations for the bilinears:

\begin{eqnarray}
\label{df}
df & = & {\textstyle\frac{1}{\sqrt3}}h_{I}i_VF^{I}\, , \\
& & \nonumber \\
\label{killingvector}                
\nabla_{(\mu}V_{\nu)} & = & 0\, ,\\
& & \nonumber \\
\label{dV}
dV & = & -{\textstyle\frac{2}{\sqrt{3}}}fh_{I}F^{I}
-{\textstyle\frac{1}{\sqrt{3}}}h_{I}{}^{\star}\left(F^{I}\wedge V\right)\, ,\\
& & \nonumber \\
\label{nablaPhi}
\mathsf{D}_{\alpha}\Phi^{r}{}_{\beta\gamma} & = &
-{\textstyle\frac{1}{\sqrt{3}}}h_{I}F^{I\,\rho\sigma}
\left(
g_{\rho[\beta}{}^{\star}\Phi^{r}{}_{\gamma]\alpha\sigma}
-g_{\rho\alpha}{}^{\star}\Phi^{r}{}_{\beta\gamma\sigma} 
-{\textstyle\frac{1}{2}}g_{\alpha[\beta}{}^{\star}
\Phi^{r}{}_{\gamma]\rho\sigma}
\right)\, ,
\end{eqnarray}

\noindent
where 

\begin{equation}
\mathsf{D}_{\alpha}\Phi^{r}{}_{\beta\gamma}=
\nabla_{\alpha}\Phi^{r}{}_{\beta\gamma} 
+2\varepsilon^{rst}\mathsf{A}^{s}{}_{\alpha} \Phi^{t}{}_{\beta\gamma}\, .  
\end{equation}

These equations are formally identical to those obtained in
Ref.~\cite{Gutowski:2004yv} but now the covariant derivative that acts on the
triplet of 2-forms is an $SU(2)$-covariant derivative.

Eqs.~(\ref{gauginokse}) and (\ref{hyperinokse}) lead to algebraic equations
for the tensor bilinears: contracting Eq.~(\ref{gauginokse}) with
$i\bar\epsilon_{i}$ and $\sigma^{r}{}_{i}{}^{j}\bar\epsilon_{j}$ we get

\begin{eqnarray}
\label{lvphi}
\pounds_{V}\phi^{x} & = & 0\, ,\\
& & \nonumber \\
\label{FtimesPhi}
h^{x}_{I}F^{I}_{\alpha\beta}\Phi^{r\,\alpha\beta} & = & 0\, ,
\end{eqnarray}

\noindent
and the contraction of Eq.~(\ref{hyperinokse}) with $i\bar\epsilon_{k}$ yields

\begin{equation}
\label{lvq}
\pounds_{V} q^{X} = 0\, .
\end{equation}

Contracting now Eq.~(\ref{gauginokse}) with $i\bar\epsilon_{i}\gamma^\mu$ and
$\sigma^{r}{}_{i}{}^{j}\bar\epsilon_{j}\gamma^{\mu}$ we get

\begin{eqnarray}
\label{dphi}
fd\phi^{x} & = & -h^{x}_{I}i_VF^{I}\, , \\
& & \nonumber \\
\label{eq:Phidphi}
0 & =& \Phi^{r}{}_{\mu\nu}\partial^{\nu}\phi^{x} 
+{\textstyle\frac{1}{4}}\varepsilon_{\mu\nu\alpha\beta\gamma} 
h^{x}_{I} F^{I\,\nu\alpha}\Phi^{r\,\beta\gamma} \, ,
\end{eqnarray}

\noindent
and, finally, operating on Eq.~(\ref{hyperinokse}) with
$\bar\epsilon_k\gamma^\mu$

\begin{equation}
\label{preholomorphicq}
f\partial_{\mu} q^{X} +\Phi^{r}{}_{\mu}{}^{\nu}\partial_{\nu} 
q^{Y} J^{r}{}_{Y}{}^{X} =0\, ,
\end{equation}

\noindent
where we have identified the complex structures of the target 
quaternionic-K\"ahler manifold,

\begin{equation}
J^{r}{}_{Y}{}^{X} = f_{Y}{}^{iA} \mathtt{J}^{r}{}_{iA}{}^{jB} f_{jB}{}^{X}\, .
\end{equation}

Eq.~(\ref{killingvector}) says that $V$ is an isometry of the space-time
metric. The differential equation of $\Phi^{r}$~(\ref{nablaPhi}) implies

\begin{equation}
\label{eq:covariantconstancy}
d\Phi^{r}+2\varepsilon^{rst}\mathsf{A}^{s}\wedge\Phi^{t}=0\, ,
\end{equation}

\noindent
i.e.~the three 2-forms are covariantly closed respect to the induced
$\mathfrak{su}(2)$ connection.

In order to make further progress, it is necessary to separate the timelike
($f\neq 0$) and null ($f=0$) cases.


\subsection{The timelike case}
\label{sec-timelike}


\subsubsection{The equations for the bilinears}

In this case the Killing vector $V$ is a timelike, $V^{2}=f^{2}>0$. We
introduce an adapted time coordinate $t$: $V = \partial_{t}$. With this choice
of coordinates the metric can be decomposed in the following way

\begin{equation}
\label{conforma-stationary}
ds^{2} = f^{2}\left(dt+\omega\right)^{2}
-f^{-1}h_{\underline{m}\underline{n}} dx^{m}dx^{n}\, ,
\end{equation}

\noindent
where $\omega$ is a time-independent 1-form and
$h_{\underline{m}\underline{n}}$ is a time-independent Riemannian
four-dimensional metric.\footnote{ Appendix~\ref{app-metric} contains a
  Vielbein basis and the non-vanishing components of the connection and Ricci
  tensor in that basis.}  Eqs.~(\ref{df}),(\ref{lvphi}) and (\ref{lvq}) imply
that with our choice of coordinates the scalars $f$, $\phi^{x}$ and $q^{X}$
are time-independent.

Following Ref.~\cite{Gauntlett:2002nw} we define the following decomposition

\begin{equation}
fd\omega = G^{+}+G^{-}\, ,
\end{equation}

\noindent
where $G^{+}$ and $G^{-}$ are the selfdual and anti-selfdual parts respect
to the metric $h$.

The Fierz identity Eq.~(\ref{VPhi}) indicates that the $\Phi^{r}$s have no
time components and the Fierz identity Eq.~(\ref{VStarPhi}) implies that they
are anti-selfdual respect to the spatial metric $h$.  Moreover, the identity
Eq.~(\ref{quaternions}) becomes

\begin{equation}
\Phi^{r}{}_{m}{}^{n} \Phi^{s}{}_{n}{}^{p} = -\delta^{rs}\delta_{m}{}^{p}
+\varepsilon^{rst}\Phi^{t}{}_{m}{}^{p}\, ,
\end{equation}

\noindent
where all operations on the spatial indices refers to the spatial metric $h$.
This is the algebra of the imaginary unit quaternions, whence we may conclude
that the spatial manifold is endowed with an \emph{almost} quaternionic
structure.

The next step is to obtain the form of the supersymmetric vector fields from
Eqs.~(\ref{df}), (\ref{dV}), (\ref{FtimesPhi}) and (\ref{dphi}): these
equations contain no explicit contributions from the hyperscalars and,
therefore lead to the same form of the vector fields found in
Ref.~\cite{Gutowski:2004yv}, namely



\begin{equation}
\label{F}
F^{I} = -\sqrt{3}\{d\left[fh^{I}\left(dt+\omega\right)\right]
+\Theta^{I}\}\, , 
\end{equation}

\noindent
where the $\Theta^{I}$s are spatial selfdual 2-forms and 

\begin{equation}
\label{remanentconstraint}
G^{+} = -{\textstyle\frac{3}{2}} h_{I}\Theta^{I}\, .
\end{equation}

From (\ref{nablaPhi}) information about the derivatives of the two-forms
$\Phi^{r}$ can be extracted using the above expression for $F^{I}$: first, by
introducing the spin connection of the metric given in
Appendix~\ref{app-metric} we may obtain the spatial components of the
five-dimensional covariant derivative,

\begin{equation}
\label{nabladecomposition}
\nabla^{(5)}_{m}\Phi^{r}{}_{nq} =  f^{3/2}\nabla_{m}\Phi_{nq}
-{\textstyle\frac{2}{3}}\left(\delta_{m[n}\partial_{p]}f^{3/2}\Phi^{r}{}_{pq}
-\delta_{m[q}\partial_{p]}f^{3/2}\Phi^{r}{}_{pn}
-\partial_{m}f^{3/2}\Phi^{r}{}_{nq}\right)\, ,
\end{equation}

\noindent
where $\nabla_{m}$ is the covariant derivative of the four-dimensional spatial
metric. On the right hand side of this expression all of the flat indices
refers to the Vielbein $v_{m}{}^{\underline{i}}$. On the other hand, the
spatial components of the equation~(\ref{nablaPhi}) are

\begin{equation}
\label{spatialnablaPhi}
\nabla^{(5)}_m\Phi^{r}{}_{nq} 
+2f^{3/2}\varepsilon^{rst} \mathsf{A}^{s}{}_{m} \Phi^{t}{}_{nq}=
-{\textstyle\frac{1}{\sqrt{3}}}fh_{I}F^{I\,p0}
\left(
\delta_{p[n}\Phi^{r}{}_{q]m}-\delta_{pm}\Phi^{r}{}_{nq}
-\delta_{m[n}\Phi^{r}{}_{q]p}
\right)
\end{equation}

\noindent
where we have used the fact that $\Phi^{r}$ are spatial, anti-selfdual 
2-forms. Now from Eq.~(\ref{F}) we read

\begin{equation}
h_{I}F^{I\,p0} = \sqrt{3}f^{-1/2}\partial_pf
\end{equation}

\noindent
and by comparing Eqs.~(\ref{nabladecomposition}) and (\ref{spatialnablaPhi})
we find that the 2-forms $\Phi^{r}$ are $SU(2)$- and Lorentz-covariantly
constant over the 4-dimensional spatial manifold:

\begin{equation}
\label{constantJ}
\nabla_{m}\Phi^{r}{}_{np}
+2\varepsilon^{rst} \mathsf{A}^{s}{}_{m} \Phi^{t}{}_{np} \ =\
\partial_{m}\Phi^{r}{}_{np} -2\xi_{m[n|}{}^{q}\Phi^{r}{}_{q|p]}
+2\varepsilon^{rst} \mathsf{A}^{s}{}_{m} \Phi^{t}{}_{np} \ =\
0\, ,
\end{equation}

\noindent
Here $\xi$ is the standard spin connection of the 4-dimensional spatial
manifold.

Had the base space not been 4-dimensional, the conclusion would have been
that we are dealing with a quaternionic-K\"ahler manifold. But in four
dimensions the above equation, taken at face value, is rather void: given a
Vierbein we can construct a kosher quaternionic structure by inducing the one
from $\mathbb{R}^{4}$ and then the unique $\mathsf{A}$ solving Eq.
(\ref{constantJ}), is given by

\begin{equation}
\label{eq:stommekutzooi}
\mathsf{A}^{r}_{m} \; =\; \textstyle{1\over 16}\ \varepsilon^{rts}\
                          \Phi^{t}_{p}{}^{n}\ \nabla_{m}\Phi^{s}_{n}{}^{p} \; .
\end{equation}

In the case at hand, however, said arbitrariness is nothing but an illusion
since the connection $\mathsf{A}$ is the one induced from an
$\mathfrak{sp}(1)$ connection on a quaternionic-K\"ahler manifold and is
therefore not to be chosen but to be deduced.  At this point one can then
already appreciate the interwoven nature of the problem: Since the
quaternionic structure on the 4-dimensional space is basically known,
Eq.~(\ref{constantJ}) determines, part of, the spin connection in terms of the
pull-back of an $\mathfrak{sp}(1)$ connection. This pull-back, however, is
defined by means of a harmonic map satisfying Eq. (\ref{preholomorphicq}),
which presupposes knowing the Vierbein, and hence also the spin connection.

A `trivial' solution to the requirement that the hyperscalars form a harmonic
map satisfying Eq. (\ref{preholomorphicq}), is to take them to be constant:
Eq. (\ref{constantJ}) then states that $\Phi$ defines a covariantly constant
hypercomplex structure, so that the 4-dimensional manifold has to be
hyper-K\"ahler, and we recover the results of
\cite{Gauntlett:2002nw,Gutowski:2004yv}.  As is well-known the holonomy of a
4-dimensional hyper-K\"ahler space is $\mathfrak{su}(2)\subset
\mathfrak{so}(4)$, and in a suitable frame the spin connection can be taken to
be selfdual. The technical reason why the spin connection can be taken to be
selfdual lies in the fact that the $\Phi$s are anti-selfdual and that the
split into anti- and selfdual components corresponds to the Lie algebraic
split $\mathfrak{so}(4)\ \cong\ 
\mathfrak{su}(2)_{+}\oplus\mathfrak{su}(2)_{-}$; if we then take the $\Phi$s
to be induced from the ones on $\mathbb{R}^{4}$, called $\mathsf{J}$, and
denote the projection of the spin connection onto $\mathfrak{su}(2)_{\pm}$ by
$\xi^{\pm}$, then Eq (\ref{constantJ}) can be expressed as $[
\xi^{-}_{m},\mathsf{J}^{r} ] = 0 $, which immediately implies $\xi^{-}=0$.

In the general case there will still be no constraint on $\xi^{+}$, but we
can solve equation (\ref{constantJ}) to give

\begin{equation}
\label{eq:embedding}
      \xi^{-}_{m\ n}{}^{q} \; =\;  -\vec{\mathsf{A}}_{m}\ \cdot\ \vec{\mathsf{J}}_{n}{}^{q} \; ,
\end{equation}

\noindent
where as above, we made use of the quaternionic structure induced from flat
space.

In the above we were able to match things up without much ado, since the
relevant $\mathfrak{su}(2)$s both acted in the vector representation. When
considering the Killing spinor equation, however, the representations do not
add up that nicely, and one finds that a necessary condition for having
unbroken supersymmetry is that the generators of $\mathfrak{su}(2)$ and
$\mathfrak{su}(2)_{-}$ should have identical actions on the Killing spinors,
{\em i.e.\/}

\begin{equation}
\epsilon^{j}\ i\sigma^{r}{}_{j}{}^{i} \; =\; 
{\textstyle\frac{1}{4}}\ \mathsf{J}^{r}{}_{mn}\gamma^{mn}\ \epsilon^{i}\, ,
\end{equation}

\noindent
and these conditions will appear as projectors $\Pi^{r\, +}{}_{i}{}^{j}$
acting on the Killing spinors, where

\begin{equation}
\label{eq:Pirpm}
\Pi^{r\, \pm}{}_{i}{}^{j}\; =\; {\textstyle\frac{1}{2}}\left[\ \delta
\ \pm\ {\textstyle\frac{i}{4}}
\not\!\mathsf{J}^{(r)}\sigma^{(r)}\right]_{i}{}^{j} \, .  
\end{equation}

\noindent
In principle we only need to impose one such constraint for each non-trivial
component $\mathsf{A}^{r}$.

The last constraint on the bosonic fields comes from
Eq.~(\ref{preholomorphicq}). In the timelike case this equation is purely
spatial and in 4-dimensional notation reads

\begin{equation}
\label{triholomorphic}
\partial_{m} q^{X} \; =\; 
\Phi^{r}{}_{m}{}^{n}\ \partial_{n} q^{Y}\ J^{r}{}_{Y}{}^{X}\, .
\end{equation}

This condition implies that $q$ is what Ref.~\cite{Chen:2000} calls a
\textit{quaternionic map}. In said reference it is shown that a quaternionic
map between hyper-K\"{a}hler manifolds implies that the map is harmonic,
\textit{i.e.}  it solves

\begin{equation}
\mathfrak{D}_{\mu}\partial^{\mu}\ q^{X} =0\, .
\end{equation}

\noindent
Here, however, we are not dealing with maps between hyper-K\"{a}hler manifolds,
yet the KSIs state that $q$ is automatically harmonic. The question then is:
Apart from being quaternionic, what properties must $q$ satisfy in order to be
harmonic?

Let us be a bit more general and consider the situation in which the
$\mathfrak{sp}(1)$ connection $\mathsf{A}$ appearing in Eq.~(\ref{constantJ})
is \textit{not} the pull-back of the $\mathfrak{sp}(1)$ connection, denoted
$\mathsf{B}$, defined on the hypervariety. By then differentiating
Eq.~(\ref{triholomorphic}), using Eqs.~(\ref{constantJ}) and the formulas in
App.~(\ref{sec-quaternionic}), we obtain

\begin{equation}
  \label{eq:KutVent}
  \begin{array}{lcl}
\mathfrak{D}_{m}\partial_{n}q^{X} 
& =&
-2\varepsilon^{str}\left[\mathsf{A}^{s}{}_{n} 
\ -\ \partial_{n}q^{Z}\ \mathsf{B}^{s}{}_{Z}
\right]\ \Phi^{t}{}_{m}{}^{p}\partial_{p}q^{Y}\ J^{r}{}_{Y}{}^{X}\\
& & \\
& & 
+\Phi^{r}{}_{n}{}^{p}\ \mathfrak{D}_{m}\partial_{p}q^{y}\
\vec{J}^{r}{}_{Y}{}^{X} \; .
  \end{array}
\end{equation}

\noindent
Contracting the free indices, we find that

\begin{equation}
\label{eq:KutVent2}
\mathfrak{D}_{m}\partial^{m}q^{X} =
2\varepsilon^{str}
\left[
\mathsf{A}^{s}{}_{m} - \partial_{m}q^{Z}\ \mathsf{B}^{s}{}_{Z}
\right] 
\Phi^{t\ nm}\partial_{n}q^{Y} J^{r}{}_{Y}{}^{X} \; .
\end{equation}

In our case, we have $\mathsf{A} = dq\cdot \mathsf{B}$ whence the fact that
$q$ is a quaternionic map, by itself, implies that it is harmonic.

Therefore, supersymmetric configurations of the hyperscalars consist of
quaternionic maps $q$ such that the $\mathfrak{su}(2)_{-}$ connection
of the 4-dimensional space manifold is canceled by
the pullback of the one on the hypervariety.

In the next section we are going to check whether the conditions that we have
derived on the fields are sufficient to have unbroken supersymmetry, 
{\em i.e.\/} identically solve the KSEs.


\subsubsection{Solving the Killing spinor equations}

We begin with Eq.~(\ref{gauginokse}), from the gaugino supersymmetry
transformation. After use of the expression of the vectorial fields
Eq.~(\ref{F}), it can be put in the form

\begin{equation}
\left(2\not\!\partial\phi^{x}
-{\textstyle\frac{\sqrt{3}}{2}}\not\!\Theta^{I}\right)
R^{-}\epsilon^{i}=0\, ,
\end{equation}

\noindent
where we have defined the projectors $R^{\pm}$

\begin{equation}
R^{\pm} \equiv {\textstyle\frac{1}{2}} \left(1\pm\gamma^{0}\right)\, .
\end{equation}

\noindent
Obviously, this equation can always be solved by imposing  the projection

\begin{equation}
\label{projection1}
R^{-}\epsilon^{i}=0\, ,  
\end{equation}

\noindent
which is equivalent to a chirality condition on the spinors over the spatial
manifold due to the relation $\gamma^{0}=\gamma^{1234}$. $R^{+}$ and $R^{-}$
have rank $2$ and therefore this projection breaks/preserves $1/2$ of the
original supersymmetries.

Now we analyze Eq.~(\ref{hyperinokse}), from the hyperinos supersymmetry
transformations. Using Eq.~(\ref{triholomorphic}) we can rewrite it in the
form

\begin{equation}
\label{eq:zooitje1}
f_{X}{}^{jA}\not\!\partial q^{X} 
\left[
3\delta_{j}{}^{i} +{\textstyle\frac{i}{4}}\sum_{r} 
\not\!\mathsf{J}^{(r)}\sigma^{(r)}{}_{j}{}^{i}\gamma_{0}
\right]\epsilon_{i} -\gamma_{m}\mathsf{J}^{r}{}_{mn}\partial_{n}q^{Y}J^{r}{}_{Y}{}^{X}
f_{X}{}^{iA}R^{-}\epsilon_{i}= 0\, ,
\end{equation}

\noindent
which can be solved by imposing the projection Eq.~(\ref{projection1}) and 

\begin{equation}
\label{projection2}
\Pi^{r+}{}_{j}{}^{i} \epsilon^{j} =0\, ,
\end{equation}

\noindent
where the $\Pi^{r\pm}{}_{j}{}^{i}$s are the objects defined in
Eq.~(\ref{eq:Pirpm}).  The $\Pi^{r+}{}_{j}{}^{i}$ satisfy the algebra

\begin{equation}
\Pi^{r\, +}\Pi^{s\, +} \; =\;  {\textstyle\frac{1}{2}}\Pi^{r\, +}  
\ +\ {\textstyle\frac{1}{2}}\Pi^{s\, +} 
\ -\ {\textstyle\frac{1}{2}}\varepsilon^{rst}\Pi^{t\, +}
\ -\ \textstyle{1\over 4}\ \delta^{rs}R^{-}\, ,
\end{equation}

\noindent
and are idempotent (and, therefore, projectors) only in the
subspace of spinors satisfying the projection Eq.~(\ref{projection1}).

Observe that, in principle, we need to impose the three projections $r=1,2,3$
on the Killing spinors. The above algebra shows that only two of them are
independent and it is easy to see that they preserve only $1/4$ of the
supersymmetries preserved by the projection Eq.~(\ref{projection1}), {\em
  i.e.} only $1/8$ of the supersymmetries is generically preserved in presence
of non-trivial hyperscalars.

We turn now to Eq.~(\ref{gravitinokse}) from the gravitino supersymmetry
transformation. We consider separately the timelike and spacelike components
of this equation.  By using the spin connection of the five-dimensional metric
Eqs.~(\ref{eq:conformaspincon}) and the expression of the vector
fields Eq.~(\ref{F}), the timelike component takes the form

\begin{equation}
\partial_{0}\epsilon^{i}
+\left[2\not\!\partial f^{1/2}
-{\textstyle\frac{1}{4}}f\left(1-{\textstyle\frac{1}{3}}\gamma^{0}\right)
\not\!G^{+}
-{\textstyle\frac{1}{4}}f\not\!G^{-}\right]
R^{-}\epsilon^{i}=0\, ,
\end{equation}

\noindent
which is automatically solved by time-independent Killing spinors satisfying
the projection Eq.~(\ref{projection1}).

The space-like components of Eq.~(\ref{gravitinokse}) take, after use of
Eq.~(\ref{projection1}), the form

\begin{equation}
\label{fingravitino}
\nabla_{m}\eta^{i}\ +\ \eta^{j} \mathsf{A}_{m\,j}{}^{i}\; =\ 0\, ,
\hspace{1cm}
\eta^{i}\equiv f^{-1/2}\epsilon^{i}\, .
\end{equation}

To solve this equation, the quaternionic nature of the 4-dimensional spatial
manifold comes to our rescue: in the special Vierbein basis and $SU(2)$ gauge
in which Eq.~(\ref{eq:embedding}) holds, the 2-forms $\Phi^{r}{}_{mn}$ are the
constants $\mathtt{J}^{r}{}_{mn}$.  Using this splitting, the above equation
takes the form

\begin{equation}
\nabla^{+}_{m}\eta^{i} \ +\  i\  \mathsf{A}^{r}{}_{m}
\left(\sigma^{r}{}_{j}{}^{i}
+{\textstyle\frac{i}{4}}
\not\!\mathsf{J}^{r}\delta_{j}{}^{i}\right)\eta^{j} =0\, ,
\hspace{1cm}
\nabla^{+}_{m}\eta^{i}\ =\
(\partial_{m}\ +\ {\textstyle\frac{1}{4}}\not\!\xi^{+}{}_{m})\eta^{i}\, .
\end{equation}

\noindent
Using the projections Eq.~(\ref{projection2}) for each non-vanishing component
of the pull-back of the $\mathfrak{su}(2)$ connection
$\mathsf{A}^{r}{}_{X}\partial_{m}q^{X}$ we are left with

\begin{equation}
\nabla^{+}_{m}\eta^{i}=0\, ,
\end{equation}

\noindent
which is solved by constant spinors that satisfy the projection
Eq.~(\ref{projection1}), {\em i.e.\/} if they are chiral in the 4-dimensional
spaces of constant time.

It should be clear from the discussion of the gravitino variations, that, for
some configurations, not all of the projections $\Pi$ need be imposed, {\em
  e.g.\/} when turning on only an $\mathfrak{u}(1)$ in $\mathfrak{su}(2)_{-}$.
The analysis of Eq.~(\ref{eq:zooitje1}), however, indicates that still all 3
of the projections ought to be implemented.  This is true if we disregard the
possibility of a special coordinate dependency of the quaternionic map. As an
extreme example we have the case with frozen hyperscalars which effectively is
like not having them at all.  A less-trivial example to this effect is
fostered by the trivial uplift of the {\em c-mapped cosmic string} analyzed in
\cite[Sec. (4.4)]{Hubscher:2006mr}, in which case the map is
holomorphic.\footnote{ In fact, part of Chen and Li's article \cite{Chen:2000}
  consists of showing that there are quaternionic maps between hyper-K\"ahler
  manifolds that are {\em not} holomorphic w.r.t.~some complex structure.}


\subsubsection{Supersymmetric solutions}

In Section~\ref{sec-ksis} we proved that timelike supersymmetric
configurations solve all the equations of motions if they solve the Maxwell
equations and Bianchi identities which we rewrite here in differential form
language for convenience:

\begin{eqnarray}
4{}^{\star}\mathcal{E}_{I} & = & 
 -d{}^{\star}\left(a_{IJ}F^{J}\right)
+{\textstyle\frac{1}{\sqrt{3}}}C_{IJK}F^{J}\wedge F^{K}\, ,\\
& & \nonumber \\
\mathcal{B}^{I} & = & dF^{I}\, .
\end{eqnarray}

We may evaluate these expressions for supersymmetric configurations using the
formula~(\ref{F}). The result is

\begin{eqnarray}
\label{wholemaxwell}
\mathcal{E}_{I}^{0} & = & 
-\frac{\sqrt{3}}{2}f^{2}\left[\nabla^{2}_{(4)}\left(h_{I}/f\right)
-{\textstyle\frac{1}{4}}C_{IJK}\Theta^{J}\cdot\Theta^{K}\right]\, , \\ 
& & \nonumber \\
\mathcal{E}_{I}^{m} & = & -2\sqrt{3}f^{3/2}C_{IJK}h^{J}
(\star_{(4)}d\Theta^{K})^{m}\, , \\
& & \nonumber \\
\left(\star\mathcal{B}^{I}\right){}^{0m} & = & 
-\sqrt{3}f^{3/2} (\star_{(4)}d\Theta^{I})^{m}\, .
\end{eqnarray}

\noindent
where, as usual, all the objects in the r.h.s. of the equations are written in
terms of the 4-dimensional spatial metric $h$.  The components
$\left(\star_{(4)}\mathcal{B}^{I}\right)^{mn}$ vanish identically, and it is
immediate to see that the KSI Eq.~(\ref{eq:ksi9}) is satisfied.

Then, the supersymmetric solutions have to satisfy only these two equations:

\begin{eqnarray}
\label{eq:hifequation}
\nabla^{2}_{(4)}\left(h_{I}/f\right)
-{\textstyle\frac{1}{4}}C_{IJK}\Theta^{J}\cdot\Theta^{K} & = & 0\, ,\\
& & \nonumber \\
d\Theta^{I} & = & 0\, ,
\end{eqnarray}

\noindent
which are identical to those found in Ref.~\cite{Gutowski:2004yv} in absence
of hypermultiplets, the difference being the quaternionic nature of the
4-dimensional space.


\subsection{Some explicit examples}
\label{sec:TLEx}

In the recent paper Ref.~\cite{Jong:2006za} Jong, Kaya and Sezgin gave
an explicit example with non-trivial and not-obviously-holomorphic
hyperscalars taking values in the symmetric space $H_{4} = SO(4,1)/SO(4)$. In
this section we are going to use the same set-up to find 5-dimensional
supersymmetric solutions and discuss the possible gravitational effects.

The four coordinates of the target are denoted by $q^{\underline X}$,
$\underline{X}=1,\ldots,4$, and take the metric on the hypervariety to be

\begin{equation}
        g_{\underline{X}\ \underline{Y}} = \Lambda^2\delta_{\underline{XY}}\,,
        \hspace{10mm}
        \Lambda(q^2) = \frac{1}{1\ -\ q^2}\,,
        \hspace{10mm}
        q^2 \equiv q^{\underline X} q^{\underline X} \leq 1 \, .
\end{equation}

\noindent
As one might have suspected this metric is Einstein, and since the
space is conformally flat, it is also trivially selfdual, meaning that we are
really dealing with an authentic 4-dimensional quaternionic-K\"ahler manifold.


A Vierbein for this metric is

\begin{equation}
        E^X \ =\ \Lambda \delta^X{}_{\underline Y}\ dq^{\underline Y}\, ,
        \hspace{1cm}
        E_X \ =\ \Lambda^{-1} \delta_X{}^{\underline Y}
                \frac{\partial}{\partial q^{\underline Y}}\, .
\end{equation}

In both the coordinate and the Vierbein basis the three complex structures are
given by the 't Hooft symbols $\rho^r{}_{XY}(=\mathsf{J}^{r}_{XY})$, which are real, constant and
antisymmetric matrices in the $X,Y$ indices.  Moreover they are
anti-selfdual\footnote{They can be seen as the three anti-selfdual
  combinations of generators of $\mathfrak{so}(4)$, {\em i.e.\/} the generators of the
  $\mathfrak{su}(2)_{-}$ subalgebra.}  and satisfy

\begin{eqnarray}
\rho^r{}_{XY}\ \rho^s{}_{YZ} & =& -\delta^{rs}\ \delta_{XZ}
\ +\ \epsilon^{rst}\ \rho^t{}_{XZ}\, , \\
 & & \nonumber \\
\rho^r{}_{XY}\ \rho^r{}_{WZ} & =& \delta_{XW}\ \delta_{YZ} 
- \delta_{XZ}\ \delta_{YW}
\ -\ \epsilon_{XYWZ}\, .
\end{eqnarray}

The anti-selfdual part of the spin connection is

\begin{equation}
\omega^{-XY} = 
2\ \left(q^{[X} E^{Y]}-{\textstyle\frac{1}{2}}\epsilon^{XYWZ} 
q^W E^Z\right)\,,
\end{equation}

\noindent
where $q^X \equiv \delta^X{}_{\underline Y}\ q^{\underline Y}$.

In order to construct the hyperscalars, we assume that also the base manifold
is conformally flat, {\em i.e.\/}

\begin{equation}
        h_{\underline{mn}}dx^{\underline m} dx^{\underline n}
                = \Omega^2 dx^{\underline m} dx^{\underline m}\,,
        \hspace{10mm}
        \Omega = \Omega(x^2)\, ,
        \hspace{10mm}
        x^2 \equiv x^{\underline m} x^{\underline m} \; ,
\end{equation}

\noindent
and thence take the Vierbein on the base manifold to be

\begin{equation}
        V^m = \Omega \delta^m{}_{\underline m}dx^{\underline m}\,,
        \hspace{10mm}
        V_m = \Omega^{-1}\delta_m{}^{\underline m}\partial_{\underline m}\,.
\end{equation}

In this basis we can identify the complex structures of the base manifold with
those of the hypervariety

\begin{equation}
        \textsf{J}^r{}_{m}{}^{n}
        \ =\ \delta_m{}^X J^r{}_X{}^Y \delta^n_Y
        \ =\ \rho^r{}_{mn}\, .
\end{equation}

\noindent
The anti-selfdual part of the spin connection on the base manifold is

\begin{equation}
        \xi^{-mn} = 2\frac{\Omega'}{\Omega^2}
\left(
x^{[m}V^{n]} - {\textstyle\frac{1}{2}}\epsilon^{mnpq} x^pV^q
\right)
\end{equation}

\noindent
where $x^m = \delta^m{}_{\underline m}\ x^{\underline n}$.

Now we analyze the conditions for supersymmetry on the hyperscalars
$q^{\underline X}$.  The first condition is that they must constitute a
quaternionic map, {\em i.e.\/} Eq. (\ref{triholomorphic}), w.r.t.~the chosen
quaternionic structures.  In our setting this equations takes the form

\begin{equation}
        \partial_{\underline m} q^{\underline X} =
        \left(\delta_{\underline{mY}}\delta_{\underline{nX}}
        - \delta_{\underline{mX}}\delta_{\underline{nY}}
        -\epsilon_{\underline{mnYX}}\right)
        \partial_{\underline n} q^{\underline Y}
\end{equation}

\noindent
whose symmetric and antisymmetric parts give

\begin{eqnarray}
\label{eq:symmquaternionic}
\partial_{\underline m}q^{\underline m} & =& 0\, , \\
\label{eq:antisymmquaternionic}
\partial_{[\underline m} q_{\underline n]} & =&
        -{\textstyle\frac{1}{2}}\epsilon_{\underline{mnpq}}
        \partial_{\underline p}q_{\underline q}\,,
\end{eqnarray}

\noindent
where $q_{\underline m} = q^{\underline m}$.

A solution to these equations is

\begin{equation}
        q^{\underline m} \; =\; x^{\underline m}\; x^{-4}\, ,
\end{equation}

\noindent
where we have chosen a possible multiplicative constant to be unity.

The second condition on the hyperscalars states that the anti-selfdual part
of the spin connection of the base manifold must be related to the
$\mathfrak{su}(2)$ connection induced from the target,

\begin{eqnarray}
&& \label{eq:relatedbaseconnections}
        \xi^-_{mn}{}^p \; =\;  -\vec{\textsf{A}}_m\cdot \vec{\textsf{J}}_n{}^p\,,
\\ &&
        \vec{\textsf{A}}_m \; \equiv\;  \partial_m q^{\underline X}\,  \vec{\omega}_{\underline X}\,,
\end{eqnarray}

\noindent
where $\vec{\omega}_{\underline X}$ is the $\mathfrak{su}(2)$ connection of
the target.  We observe that the reasoning leading to the
relation~(\ref{eq:relatedbaseconnections}) can be applied on the target
manifold as well,\footnote{ Indeed it can be applied in any four-dimensional
  Riemannian manifold.  }, where the involved connections are $\omega_{XY}$
and $\vec{\omega}_{\underline X}$ and therefore we may establish the
following relation on the target

\begin{equation}
\label{eq:relatedtargetconnections}
        \omega^-_{XY}{}^Z \; =\;  -\vec{\omega}_X\  \cdot\ \vec{J}_Y{}^Z\,.
\end{equation}

\noindent
By contrasting
Eqs.~(\ref{eq:relatedbaseconnections})-(\ref{eq:relatedtargetconnections}) we
conclude that in our settings the anti-selfdual part of the spin connection
of the base manifold is induced from the one of the hypervariety,

\begin{equation}
        \xi^-_{m}{}^{np} \; =\;
        \partial_m q^{\underline X}
        \omega^{-}_{\underline{X}}{}^{YZ} \delta_{YZ}{}^{np}\,.
\end{equation}

\noindent
This condition is satisfied if

\begin{equation}
        \frac{\Omega'}{\Omega} = \frac{ 1}
        {x^2\left(x^6 - 1\right)}\,.
\end{equation}

\noindent
The solution to this equation is

\begin{equation}
\label{eq:Omega}
        \Omega = \left(1\ -\ x^{-6}\right)^{1/3}\,,
\end{equation}

\noindent
where, as above, we chose a certain multiplicative integration constant.  We
would like to point out that in this case the whole spin connection on the
base manifold, rather than only its anti-selfdual part, is induced by the
connection on the hypervariety.

A small investigation of the curvature invariants for the metric on the base space,
shows that the point $x^{2} = 1$ corresponds to a naked curvature singularity.


We have, thus, found the following $1/8$ BPS, static, asymptotically flat,
spherically symmetric, solution with only unfrozen hyperscalars in the
$SO(1,4)/SO(4)$ coset:

\begin{equation}
\begin{array}{rcl}
ds^2 & =&  dt^{2} -{\displaystyle\left(1-\frac{1}{x^6}\right)^{2/3}}
dx^{\underline m} dx^{\underline m}\, , \\
& & \\
q^{\underline m} & = & {\displaystyle\frac{x^{\underline m}}{x^{4}}}\, ,\\
\end{array}
\end{equation}

which, as was said above, presents a naked singularity at $x^{2}=1$. Since there are no
conserved charges in this system, the \textit{no hair} conjecture suggests
that black-hole type (i.e.~spherically symmetric) solutions of this and
similar systems will always be singular, but a more detailed study is needed
to reach a final conclusion since they may be excluded by a mechanism like the
one discussed in Ref.~\cite{Bellorin:2006xr,kn:BMO}. Furthermore, a
higher-dimensional stringy interpretation of this, and similar solutions, is also
needed as to interpret this singularity correctly.

%
As a further example let us now consider how solutions of minimal $N=1,d=5$
SUGRA\footnote{In our notation this means that $n_{v}=0$, $C_{111}=1$ and
  $h^{1}=1$.} are deformed by the coupling to these hyperscalars.  For the
sake of simplicity we consider the simplest static ($\Theta=\omega=0$)
solution which is determined, according to Eq.~(\ref{eq:hifequation}), by a
single function $f^{-1}=K$ which is harmonic w.r.t.~the metric on the base
manifold. The supersymmetric solution can be written as

\begin{equation}
  \begin{array}{rcl}
ds^2 & = & K^{-2}\ dt^2 \ -\ K\ 
{\displaystyle\left(1+\frac{\lambda}{x^6}\right)^{2/3}}\
dx^{\underline m} dx^{\underline m}\, , \\
& & \\
A & = & -\sqrt{3}\ K^{-1}\  dt\, , \\
& & \\
q^{\underline m} & = & {\displaystyle\frac{x^{\underline m}}{x^4}}\,.
\end{array}
\end{equation}

If the harmonic function is chosen as to have an asymptotically flat,
spherically symmetric solution with positive mass, the harmonic function is,
with frozen hyperscalars,

\begin{equation}
K= 1 +\frac{|Q|}{x^{2}}\, ,
\end{equation}

\noindent
and the solution is the 5-dimensional Reissner-Nordstr\"om black hole
\cite{Myers:1986un} which has an event horizon at $x=0$ that covers all
singularities.

When the  hyperscalars are unfrozen and we have the above base manifold,
$K$, determined again by imposing asymptotic flatness and spherical symmetry,
is given by 

\begin{equation}
        K \; =\; 1\ +\ Q\ \frac{\mathrm{_{2}F_{1}}
        \left({\textstyle\frac{1}{3},\frac{2}{3}}\ ;\
        {\textstyle\frac{4}{3}}\ ;\ x^{-6}\right)}{x^{2}} \; ,
\end{equation}

\noindent 
where $_{2}F_{1}$ is a Gau{\ss} hypergeometric function.  It is easy to see
that $\lim_{x^{2}\rightarrow \infty} K \ =\ 1$ and that $ \mathrm{_{2}F_{1}}
\left({\textstyle\frac{1}{3},\frac{2}{3}}\ ;\ {\textstyle\frac{4}{3}}\ ;\ 
  x^{-6}\right)/x^{2}$ is a real, strictly positive and monotonically decreasing function
on the interval $x^{2}\in (1,\infty)$.  
The real question then is:
what happens at $x^{2}=1$?  Eq.~\cite[15.1.20]{abramowitz} gives a
straightforward answer

\begin{equation}
  _{2}F_{1}\left(\
        \textstyle{1\over 3},\textstyle{2\over 3};\textstyle{4\over 3};1\
    \right) \; =\; \frac{
            \Gamma\left(\textstyle{1\over 3}\right)
            \Gamma\left(\textstyle{4\over 3}\right)
}{
            \Gamma\left(\textstyle{2\over 3}\right)
}\;\sim\; 1.76664 \; ,
\end{equation}

\noindent
which implies that there is a naked singularity at $x^{2}=1$.




\subsubsection{Solutions with an additional isometry}
\label{sec-timelikeisometry}

To make contact with the families of solutions with one additional isometry
found in Refs.~\cite{Gauntlett:2002nw,Gauntlett:2004qy} we make the following
\textit{Ansatz} for the 4-dimensional spacelike metric

\begin{equation}
\label{eq:GGH}
h_{\underline{m}\underline{n}}dx^{m}dx^{n} =  H^{-1} (dz +\chi)^{2}
+H \gamma_{\underline{r}\underline{s}}dx^{r}dx^{s}\, ,\,\,\,\, r,s=1,2,3\, ,
\end{equation}

\noindent
where the function $H$, the 3-dimensional metric
$\gamma_{\underline{r}\underline{s}}$, and the 1-form
$\chi=\chi_{\underline{r}}dx^{r}$ are all independent of the coordinate $z$.
This \textit{Ansatz} covers all 4-dimensional metrics with one isometry. We
also require all fields in the solution to be independent of $z$.

As we have seen, supersymmetry requires the anti-selfdual part of the spin
connection of this metric to be identical to the pullback of the
$\mathfrak{su}(2)$ connection of the hypervariety. With the orientation
$\varepsilon_{z123}=+1$ and the Vierbein basis

\begin{equation}
V^{z} = H^{-1/2}(dz+\chi)\, ,
\hspace{1cm}
V^{r} = H^{1/2}v^{r}\, , 
\end{equation}

\noindent
where the $v^{r}$ is the Dreibein for the 3-dimensional metric
$\gamma_{\underline{r}\underline{s}}$, the anti-selfdual part of the spin
connection 1-form is given by

\begin{equation}
  \begin{array}{rcl}
\xi^{-\, zr} & = & {\textstyle\frac{1}{2}}H^{-3/2}[\partial_{r}H
-(\hat{\star}\hat{d}\chi)_{r}]V^{z}  \\
& & \\
& & 
+{\textstyle\frac{1}{4}}\varepsilon_{rst}H^{-3/2}
\{
[\partial_{t}H-(\hat{\star}\hat{d}\chi)_{t}]\delta_{su}
-2H\varpi_{ust}
\}V^{u}\, ,\\
\end{array}
\end{equation}

\noindent
where hatted objects refer to the 3-dimensional metric.

Observe that the $z$-independence of all fields means that the pullback of the
$\mathfrak{su}(2)$ connection has no $z$ component. Then, the supersymmetry
condition Eq.~(\ref{eq:embedding}) leads to

\begin{equation}
\label{eq:GH}
\hat{d}H = \hat{\star}\hat{d}\chi\, ,\,\,\,\, \Rightarrow\,\,\,\, 
\hat{\nabla}^{2} H=0\, ,
\end{equation}

\noindent
which is a condition on the 4-dimensional metric, and

\begin{equation}
\label{eq:embeddingiso}
\xi_{\underline{r}}^{-\, zs}
\ =\
-{\textstyle\frac{1}{2}}\varepsilon^{stu}\ \varpi_{\underline{r}}{}^{tu} 
\ =\
-2\mathsf{A}^{s}{}_{X}\ \partial_{\underline{r}}q^{X}\, ,  
\end{equation}

\noindent
which is a condition on the hyperscalars and the 3-dimensional metric.

Observe that the above 4-dimensional metric is a generalization of the
Gibbons-Hawking instanton metric \cite{Gibbons:1979zt}. The non-trivial
3-dimensional metric destroys the selfduality of the connection. However, the
generalized metric admits a quaternionic structure which is the
straightforward generalization of that of the Gibbons-Hawking metric
\cite{Gibbons:1987sp} and is, therefore, associated to the three hyper-K\"ahler
2-forms

\begin{equation}
J^{r}\equiv V^{z}\wedge V^{r}
-{\textstyle\frac{1}{2}}\varepsilon^{rst}V^{s}\wedge V^{t}\, .  
\end{equation}

\noindent
It is trivial to check that they satisfy the quaternionic algebra since the
tangent space components of these 2-forms are identical to those of the
Gibbons-Hawking metric and are proportional to the anti-selfdual generators
of $SO(4)$. Unlike the Gibbons-Hawking case, however, the hyper-K\"ahler
2-forms are not closed. Instead, a simple calculation shows that they satisfy

\begin{equation}
dJ^{r} -\varpi^{rs}\wedge J^{s}=0\, ,  
\end{equation}

\noindent
which, on account of Eq.~(\ref{eq:embeddingiso}), can be written in the form 

\begin{equation}
dJ^{r} + 2\varepsilon^{rst}\mathsf{A}^{s}\wedge J^{s}=0\, .
\end{equation}

Thus, the 4-dimensional metric Eq.~(\ref{eq:GGH}) and hyperscalars subject to
Eqs.~(\ref{eq:GH}) and (\ref{eq:embeddingiso}) (plus
Eq.~(\ref{triholomorphic})) are the most general ones associated to
supersymmetric solutions with one isometry. Using them it 
can be shown
that the general solutions found in Ref.~\cite{Gauntlett:2004qy} are
formally identical, the only difference being that the $2\bar{n}+2$ harmonic
functions $K^{I},L_{I},M,H$ on which these solutions depend, are harmonic
functions w.r.t. the 3-dimensional metric
$\gamma_{\underline{r}\underline{s}}$.

To be explicit, in terms of these harmonic functions, the scalars, the
closed selfdual 2-forms $\Theta^{I}$, and the 1-form $\omega$ take the form

\begin{equation}
  \begin{array}{rcl}
h_{I}/f & = & C_{IJK}K^{J}K^{K}/H +L_{I}\, ,\\
& & \\
\Theta^{I} & = & 
[ (dz+\chi)\wedge  d(K^{I}/H) +H\hat{\star} d(K^{I}/H)]\, ,\\
& & \\
\omega & \equiv & \omega_{5} (d\psi +\chi) +\hat{\omega}\, ,\\
& & \\
\omega_{5} & = & 
M+ {\textstyle\frac{3}{2}}H^{-1}L_{I}K^{I} 
+H^{-2}C_{IJK}K^{I}K^{J}K^{K}\, ,\\
& &  \\
\star_{(3)}d\hat{\omega} & = & 
HdM-MdH +{\textstyle\frac{3}{2}} (K^{I}dL_{I}-L_{I}dK^{I})\, .\\
\end{array}
\end{equation}

\noindent
The function $f$ has to be determined case by case using the constraint
$C_{IJK}h^{I}h^{J}h^{K}=1$, but an explicit expression for symmetric spaces is
given in Ref.~\cite{Gauntlett:2004qy}. In the $n=0$ case, {\em i.e.\/} only
one function $K^{0}\equiv K$ and one function $L_{0}\equiv L$, it is given by

\begin{equation}
f^{-1} = K^{2}/H +L\, . 
\end{equation}

The metric of these solutions can be cast in the form

\begin{equation}\label{eq:Naampje}
  \begin{array}{rcl}
ds^{2} & = & -k^{2}[dz+B]^{2} \\
& & \\
& & 
+k^{-1}
\left[
\left(\frac{fH^{-1}}{(f^{-1}H^{-1}-f^{2}\omega^{2}_{5})^{1/2}}\right)
(dt+\hat{\omega})^{2} 
-\left(\frac{fH^{-1}}{(f^{-1}H^{-1}-f^{2}\omega^{2}_{5})^{1/2}}\right)^{-1}
\gamma_{\underline{r}\underline{s}}dx^{r}dx^{s}
\right]\, ,\\
& & \\
k^{2} & = & f^{-1}H^{-1}-f^{2}\omega^{2}_{5}\, ,\\
& & \\
B & = & \chi +f^{2}\omega_{5}k^{-2}(dt+\hat{\omega})\, .\\
\end{array}
\end{equation}

\noindent
In this form, comparing with the results of
Refs.~\cite{Meessen:2006tu,Hubscher:2006mr} it is easy to see the form of the
$N=2,d=4$ supersymmetric solution that will appear after dimensional
reduction. The metric

\begin{equation}
ds^{2}=\left(\frac{fH^{-1}}{(f^{-1}H^{-1}-f^{2}\omega^{2}_{5})^{1/2}}\right)
(dt+\hat{\omega})^{2} 
-\left(\frac{fH^{-1}}{(f^{-1}H^{-1}-f^{2}\omega^{2}_{5})^{1/2}}\right)^{-1}
\gamma_{\underline{r}\underline{s}}dx^{r}dx^{s}\, ,
\end{equation}

\noindent
is that of a solution in the timelike class, to which all $N=2,d=4$
supersymmetric black holes belong, and there is an additional scalar ($k$) and
an additional vector field ($B$). If the 5-dimensional solution is static
$\omega_{5}=0$ and the vector field $B=\chi$ is magnetic and corresponds to a
KK monopole or a generalization thereof. This fact has been used in
Refs.~\cite{Bena:2004tk,Bena:2005ay,Gaiotto:2005gf,Elvang:2005sa,Gaiotto:2005xt,Bena:2005ni,Behrndt:2005he}
to relate 4- and 5-dimensional black hole solutions.


\subsection{The null case}
\label{sec-null}

Denote the null Killing vector by $l^{\mu}$. Following the same considerations
as in Refs.~\cite{Gauntlett:2002nw,Gutowski:2005id}, we find that we can
choose null coordinates $u$ and $v$ such that

\begin{equation}
l_{\mu}dx^{\mu}= f du\, ,
\hspace{1cm}
l^{\mu}\partial_{\mu} =\partial_{\underline{v}}\, ,  
\end{equation}

\noindent
where $f$ may depend on $u$ but not on $v$, and the metric can be put in the
form

\begin{equation}
\label{eq:nullmetric}
ds^{2}= 2fdu(dv+Hdu+\omega)
-f^{-2}\gamma_{\underline{r}\underline{s}}dx^{r}dx^{s}\, ,  
\end{equation}

\noindent
where $r,s,t=1,2,3$ and the 3-dimensional spatial metric
$\gamma_{\underline{r}\underline{s}}$ may also depend on $u$ but not on $v$.
Eqs.~(\ref{lvphi}) and (\ref{lvq}) state that the scalars are $v$-independent.

The above metric is completely equivalent to the one used in
Refs.~\cite{Gauntlett:2002nw,Gutowski:2005id}, but we find this form more
convenient; a Vielbein, and the corresponding spin connection and curvature
for it are given in Appendix~\ref{app-nullmetric}.

In the null case the Fierz identities (\ref{VPhi},\ref{VStarPhi}) and
(\ref{quaternions}) imply that the 2-forms bilinears $\Phi^{r}$ are of the
form

\begin{equation}
\Phi^{r} = du\wedge v^{r}\, ,  
\end{equation}

\noindent
where the 1-forms $v^{r}$ are an orthogonal basis for the 3-dimensional
spatial metric $\gamma_{\underline{r}\underline{s}}$.
Eq.~(\ref{eq:covariantconstancy}) then implies the equation

\begin{equation}
du\wedge D v^{r} =0\, ,  
\end{equation}

\noindent
{\em i.e.\/} the spatial components of the pullback of the $\mathfrak{su}(2)$
connection are related to the spin connection coefficients of the basis
$v^{r}$ (computed for constant $u$) by

\begin{equation}
\label{eq:FixOmegaNull}
\varpi_{\underline{r}}{}^{st} \; =\; 2\varepsilon^{stp}\
\mathsf{A}^{p}{}_{X}\ \partial_{\underline{r}} q^{X}\, .  
\end{equation}

\noindent
This equation is identical to the one found in Ref.~\cite{Hubscher:2006mr} in
the context of ungauged $N=2,d=4$ supergravity coupled to hypermultiplets.
Actually, substituting the 2-forms we found into Eq.~(\ref{preholomorphicq})
we arrive at

\begin{equation}
\label{eq:HypSusyNull}
\partial_{r}q^{X} f_{X}{}^{iA}\sigma^{r}{}_{i}{}^{j}=0\, ,  
\end{equation}

\noindent
which is identical to the equation that the hyperscalars have to satisfy in a
supersymmetric configuration of ungauged $N=2,d=4$ supergravity
\cite{Hubscher:2006mr}. Observe that the last two equations together with
Eq.~(\ref{eq:una}) (for $\nu=-1$) imply that the Ricci scalar of the
3-dimensional metric $\gamma$ satisfies

\begin{equation}
\label{eq:Riccisrelation}
R_{rs}(\gamma) = -{\textstyle\frac{1}{2}} g_{XY}\partial_{r}q^{X}
\partial_{s}q^{Y}\, .  
\end{equation}

Let us now determine the vector field strengths:
Eqs.~(\ref{df},\ref{FtimesPhi}) and (\ref{dphi}) lead to

\begin{equation}
l^{\mu}F^{I}_{\mu\nu}=0\, ,  
\end{equation}

\noindent
and, using the basis given in Appendix~\ref{app-nullmetric}, we can write

\begin{equation}
F^{I} = F^{I}{}_{+r}e^{+}\wedge e^{r} 
+{\textstyle\frac{1}{2}}F^{I}{}_{rs}e^{r}\wedge e^{s}=
    F^{I}{}_{+r}du\wedge v^{r} 
+{\textstyle\frac{1}{2}} f^{-2}F^{I}{}_{rs}v^{r}\wedge v^{s}\, .
\end{equation}

From Eq.~(\ref{dV}) we get\footnote{Unless stated otherwise (as is the case of
  $F^{I}{}_{rs}$) all quantities with flat spatial indices refer to the
  3-dimensional metric and Dreibein basis.}

\begin{equation}
\label{eq:FIrsNull}
h_{I}F^{I}{}_{rs} =-\sqrt{3}\varepsilon_{rst}\partial_{t}f\, , 
\hspace{1cm}
\partial_{t} \equiv v_{t}{}^{\underline{s}}\partial_{\underline{s}}\, .
\end{equation}

\noindent
The same result can be obtained from $D\star\Phi^{r}$. From
Eq.~(\ref{eq:Phidphi}) we get 

\begin{equation}
\label{eq:FxrsNull}
h^{x}_{I}F^{I}{}_{rs} \; =\; -\varepsilon_{rst} f\ \partial_{t}\phi^{x}\, ,
\end{equation}

\noindent
which, together with the previous equation and the definition of $h^{x}_{I}$
give

\begin{equation}
f^{-2}\ F^{I}{}_{rs} \; =\;  \sqrt{3}[\hat{\star}\ \hat{d}(h^{I}/f)\ ]_{rs}\, .
\end{equation}

From the $++r$ components of Eq.~(\ref{nablaPhi}) we get 

\begin{equation}
\label{eq:FIprNull}
h_{I}F^{I}{}_{+r} =-{\textstyle\frac{1}{\sqrt{3}}}f^{2}(\hat{\star}F)_{r}\, ,
\end{equation}

\noindent
where

\begin{equation}
F =\hat{d}\omega\, .
\end{equation}

The components $h^{x}_{I}F^{I}{}_{+r}$ are not determined by supersymmetry and
we parametrize them by 1-forms $\psi^{I}$ satisfying $h_{I}\psi^{I}=0$.  In
conclusion, the vector field strengths are given by

\begin{equation}
F^{I}= 
[{\textstyle\frac{1}{\sqrt{3}}} f^{2}h^{I}\hat{\star}F -\psi^{I}]\wedge du
+\sqrt{3}\hat{\star} \hat{d}(h^{I}/f)\, .
\end{equation}


\subsubsection{Solving the Killing spinor equations}
\label{sec:SolvKSENull}


Let us continue our analysis by plugging our
configuration into Eq.~(\ref{gauginokse}): using the Vielbein,
Eq.~(\ref{eq:FxrsNull}) and some Clifford algebra manipulations, we see that

\begin{equation}
  0 \, =\, f^{-1}\left[
             \partial_{u}\phi^{x} \ +\
             h^{x}_{I}\psi^{I}_{r}\ \gamma^{r} \ +\
             \textstyle{f^{2}\over 2}\partial_{t}\phi^{x}\
                \varepsilon_{trs}\gamma^{rs}\gamma^{-}
           \right]\ \gamma^{+}\epsilon^{i} \; ,
\end{equation}

\noindent
so, if we want the scalars $\phi$ and the $\psi^{I}$ to be non-trivial, we are
forced to impose $\gamma^{+}\epsilon^{i} \ =\ 0$.

As is usual in wave-like supersymmetric solutions, the $-$ component of the
susy variation (\ref{gravitinokse}) is identically satisfied by an
$v$-independent spinor, and the remainder of the components simplify greatly
due to the lightlike constraint: The ones in the $r$-directions reduce, after
using Eqs.~(\ref{eq:FIrsNull},\ref{eq:FIprNull}), to

\begin{equation}
  \label{eq:RSNullr}
  \begin{array}{lcl}
    0 & =& f\ \mathsf{D}_{r}\epsilon \; =\; 
           f\left[
                \partial_{r} \ -\
                \textstyle{1\over 4}\varpi_{rst}\gamma^{st} \ +\
                i\vec{\mathsf{A}}\cdot\vec{\sigma}^{T}
              \right]\ \epsilon \\
      & & \\
      & =& f\left[
               \partial_{r} \ +\
               \mathsf{A}_{r}^{p}\gamma^{p}\left(
                        1\ -\ i\gamma^{p}(\sigma^{(p)})^{T}
                      \right)\ 
            \right]\ \epsilon \; ,
  \end{array}
\end{equation}

\noindent
where in the last step we made use of Eq.~(\ref{eq:FixOmegaNull}).
If we then introduce the projection operators (no sum over $p$!)

\begin{equation}
  \label{eq:DefNullProj}
  \Pi_{p} \; =\; \textstyle{1\over 2}\left(
                       1\ -\ i\gamma^{p}(\sigma^{(p)})^{T}
                    \right) 
  \hspace{.5cm};\hspace{.5cm} \Pi_{p}^{2} \ =\ \Pi_{p}
  \hspace{.5cm};\hspace{.5cm} \left[\ \Pi_{p}\ ,\ \Pi_{q}\ \right] \ =\ 0\; ,
\end{equation}

\noindent
the above equation is solved by imposing the condition $\Pi_{p}\epsilon =  0$,
for every $p$ for which $\mathsf{A}^{p}$ does not vanish, leading to a Killing
spinor that can only depend on $u$.

The penultimate equation that needs to be checked is the gravitino variation
in the $u$-direction. 

\begin{equation}
  0 \, =\, \partial_{u}\epsilon \ +\ 
        \textstyle{1\over 4}v_{r}{}^{\underline{t}}\partial_{u}v_{s\underline{t}}\ 
             \gamma^{rs}\epsilon \ +\
        i\vec{\mathsf{A}}_{u}\cdot\vec{\sigma}^{T}\epsilon
    \; =\; \partial_{u}\epsilon \ -\
           \left[
              \mathsf{A}^{p}{}_{u} \ +\
              \textstyle{1\over 4}\varepsilon_{prs}
                  v_{r}{}^{\underline{t}}\partial_{u}v_{s\underline{t}}
           \right]\ \gamma^{p}\epsilon\; . 
\end{equation}

Generically the factor $v_{r}{}^{\underline{t}}\partial_{u}v_{s\underline{t}}$
is spacetime dependent, which, in order to avoid an inconsistency with
the $x$-independency of the Killing spinor, means that we must have

\begin{equation}
  \label{eq:NullAConsistency}
    \mathsf{A}^{p}{}_{u} \; =\; 
              -\textstyle{1\over 4}\varepsilon_{prs}\
                  v_{r}{}^{\underline{t}}\ \partial_{u}v_{s\underline{t}} \, .
\end{equation}

A consequence of this analysis is that the Killing spinor is constant.

Eq.~(\ref{hyperinokse}) is the only one left to be analyzed. In fact it is
straightforward to see that, given the constraints obtained thus far,
Eq.~(\ref{hyperinokse}) is tantamount to (\ref{eq:HypSusyNull}) contracted
with $\epsilon_{j}$. In order to get this far, however, one has to make use of
all the constraints, meaning that if we do not want even more constraints,
Eq.~(\ref{eq:HypSusyNull}) must hold.


\subsubsection{Equations of motion}

In the null case, the KSIs contain far less restrictive information than in
the timelike case, and as one can see from
Eqs.~(\ref{eq:ksi11})-(\ref{eq:ksi15}), there are more equations of motion to
be checked.
 
In order to get on with the show, let us analyze the gauge sector: the
non-vanishing components of the Bianchi identities are immediately found to be

\begin{eqnarray}
\label{eq:B+-}
{}^{\star}\mathcal{B}^{I\, +-} & = & 
\sqrt{3}f^{3}\hat{\nabla}^{2}(h^{I}/f)\, ,\\
& & \nonumber \\
\label{eq:B-r}
f^{-1}{}^{\star}\mathcal{B}^{I\, -r} & = & [\hat{\star}\hat{d}(
{\textstyle\frac{1}{\sqrt{3}}}f^{2}h^{I}\hat{\star}F-\psi^{I})]_{r}
+\sqrt{3}\left[\hat\star\partial_{\underline{u}}\hat{\star}\hat{d}(h^{I}/f)
\right]_r
\, ,
\end{eqnarray}

\noindent
and the Maxwell equations take the form

\begin{equation}
4{}^{\star}\mathcal{E}_{I} = -\sqrt{3} du\wedge
\left\{
f\hat{d}h_{I}\wedge F +
{\textstyle\frac{1}{\sqrt{3}}}
\left[
\hat{d}(\hat{\star}\psi_{I}/f)
-2C_{IJK}\psi^{J}\wedge \hat{\star}\hat{d}(h^{K}/f)
\right]
\right\}\, ,
\end{equation}

\noindent
and satisfy the KSIs Eqs.~(\ref{eq:ksi12}) and (\ref{eq:ksi13}).

Eq.~(\ref{eq:B+-}) is solved by $\bar{n}\equiv n_{v}+1$ harmonic\footnote{In
  this section, harmonic means harmonic on the 3-dimensional Euclidean space
  with metric $\gamma$.} functions $K^I$:

\begin{equation}
\label{eq:defKI}
        h^{I}/f \ =\ K^{I}\hspace{.4cm},\hspace{.4cm}
        \hat\nabla^{2}\ K^{I} \ =\ 0 \, ,
\end{equation}

\noindent
$K^I\neq 0$, which, as in the timelike case, determines $f$ to be

\begin{equation}
\label{eq:Nullfixf}
        f^{-3} \ =\  K_IK^I\,,\;\;\;
        K_{I}\ \equiv\  C_{IJK} K^{J}\ K^{K}\, .
\end{equation}

Since the $K^I$ are harmonic, we may introduce $\bar n$ local, 3-dimensional
1-forms $\alpha^I = \alpha^I_{\underline r}(u,\vec{x}) dx^r$ which satisfy

\begin{equation}
\label{defalpha}
        \hat d\alpha^I = \hat\star\hat d K^I \,,
\end{equation}

\noindent
such that each $\alpha^I$ is determined, up to a 3-dimensional gradient, in
terms of $K^I$ and $\gamma$.  This gauge freedom will be relevant soon.

Eqs.~(\ref{eq:B-r}) become

\begin{equation}
        \hat d\psi^I =
        {\textstyle\frac{1}{\sqrt3}}\hat d\left(f^2h^I\hat\star F\right) 
+\sqrt3\hat d\dot\alpha^I\,,
\end{equation}

\noindent
where $\dot\alpha \equiv \dot\alpha^I_{\underline r}\ dx^r$. The general,
local solution to this equation is

\begin{equation}
\label{eq:NCExpPsi}
\psi^I = 
{\textstyle\frac{1}{\sqrt3}} f^2h^I\hat\star F 
+\hat d M^I +\sqrt3\dot\alpha^I\,,
\end{equation}

\noindent
where the $M^I$s are some functions. The constraint $\textstyle{h\cdot \psi
  =0}$ implies

\begin{equation}\label{eq:NCFixOmega}
        {\textstyle\frac{1}{\sqrt3}}f^2\hat\star F + h_I\hat d M^I
        +\sqrt3 h_{I}\dot{\alpha}^{I} \ =\ 0\, .
\end{equation}

\noindent
Due to the relation $F = \hat{d}\omega$, the above is the equation that, if we
manage to fix the $M$s, will determine $\omega$.

Plugging Eq. (\ref{eq:NCExpPsi}) into the Maxwell equations we see that 

\begin{equation}
        \hat\nabla^2 L_I
        \ +\ \sqrt3 C_{IJK}\left[\hat\nabla_r\left(K^J\dot\alpha^K\right)_r
        +\partial_r K^J\left(\dot\alpha^K\right)_r\right] \; =\; 0\, ,
\end{equation}

\noindent
where we have defined the combinations

\begin{equation}
\label{defL}
        L_{I} \;\equiv\;  C_{IJK}\ K^{J}\ M^{K}\, .
\end{equation}

At this point we take advantage of the gauge freedom of (\ref{defalpha}) in
order to simplify the Maxwell equations: fix the gauge by imposing

\begin{equation}
\label{eq:fixalpha}
        C_{IJK}\left[\hat\nabla_r\left(K^J\dot\alpha^K\right)_r
        +\partial_r K^J\left(\dot\alpha^K\right)_r\right] \; =\; 0\, ,
\end{equation}

\noindent
thus determining $\alpha^I$ completely in terms of the $K^I$ and $\gamma$. 
In this gauge the functions $L_I$ are harmonic,

\begin{equation}
        \hat\nabla^{2}\ L_{I} \; =\; 0\; ,
\end{equation}

\noindent
and we determine the functions $M^I$ in terms of the harmonic functions $K^I$
and $L_I$ by Eq.~(\ref{defL}).

Another advantage of the above gauge is that the equation for $\omega$, Eq.
(\ref{eq:NCFixOmega}), takes on the rather nice form:

\begin{equation}
\label{eq:domega}
        \hat{\star}\hat{d}\omega \ =\  \sqrt{3}\left( L_{I}dK^{I} \ 
-\ K^{I}dL_{I}\right) 
 \ -\  3K_{I}\dot{\alpha}^{I}\, . 
\end{equation}

In the analysis of the Einstein equations it is useful to perform the
following change of variables

\begin{equation}
        H \ =\ -{\textstyle\frac{1}{2}}L_{I}M^{I} \ +\ N\, .
\end{equation}

\noindent
With this redefinition ${\cal E}_{++}$ becomes

\begin{eqnarray}
{\cal E}_{++} & = & 
-f\nabla^2 N+ f\left[\nabla_r(\dot\omega)_r + 3(\dot\omega)_r\partial_r\log f 
+{\textstyle\frac{1}{2}}f^{-3}(\ddot\gamma)_{rr} 
+ {\textstyle\frac{1}{4}}f^{-3}(\dot\gamma)^2
- {\textstyle\frac{3}{2}} f^{-4}\dot f(\dot\gamma)_{rr}\right. \nonumber \\ 
& & \nonumber \\
& &
-3C_{IJK} K^I \left(\dot K^J \dot K^K
+(\dot\alpha^J)_r (\dot\alpha^K)_r
+{\textstyle\frac{2}{\sqrt3}} (\dot\alpha^J)_r \partial_r M^K\right)
+12f^3\left(K_I\dot K^I\right)^2 \nonumber \\ 
& & \nonumber \\
& &
\left.+{\textstyle\frac{1}{2}}f^{-3} g_{XY}\dot q^X \dot q^Y\right]\, .
\end{eqnarray}

In general there is a gauge freedom in setting the one-form $\omega$ given in
(\ref{eq:domega}), corresponding to shifts in the coordinate $v$.  If we
choose to fix this gauge freedom by demanding

\begin{eqnarray}
 \nabla_r(\dot\omega)_r + 3(\dot\omega)_r\partial_r \log f & = & 
- {\textstyle\frac{1}{2}}f^{-3}(\ddot\gamma)_{rr}
- {\textstyle\frac{1}{4}}f^{-3}(\dot\gamma)^2\nonumber
+ {\textstyle\frac{3}{2}} f^{-4}\dot f(\dot\gamma)_{rr}
-{\textstyle\frac{1}{2}}f^{-3} g_{XY}\dot q^X \dot q^Y  \nonumber \\
& & \nonumber  \\
& &
+3C_{IJK} K^I \left(\dot K^J \dot K^K
+(\dot\alpha^J)_r (\dot\alpha^K)_r
+{\textstyle\frac{2}{\sqrt3}} (\dot\alpha^J)_r \partial_r M^K\right)
\nonumber \\
& & \nonumber \\
& & 
-12f^3\left(K_I\dot K^I\right)^2 \, ,  \label{eq:fixomega} 
\end{eqnarray}

\noindent
then ${\cal E}_{++}$ vanishes identically if $N$ is a real, harmonic function.
${\cal E}_{+r}$ becomes

\begin{equation}
\label{eq:E+r}
        {\cal E}_{+r} =
        -{\textstyle\frac{1}{2}}\nabla_s(\dot\gamma)_{rs}
        +{\textstyle\frac{1}{2}}\partial_r(\dot\gamma)_{ss}
        +{\textstyle\frac{3}{2}} f^3 \dot K_I\partial_r K^I
        +{\textstyle\frac{1}{2}}g_{XY}\dot q^X \partial_r q^Y\,,
\end{equation}

\noindent
whereas ${\cal E}_{rs}$ is identically satisfied by the configuration as we
have it.

\subsubsection{u-independent solutions}
\label{sec-nullisometry}


The equations that need to be solved, simplify greatly if we consider the
case that the solutions do not depend on the coordinate $u$:
in that case the gauge-fixings
Eqs.~(\ref{eq:fixalpha},\ref{eq:fixomega}) and the remaining equation of
motion, Eq.~(\ref{eq:E+r}), vanish identically, meaning that now the solutions
are completely determined by the hyperscalars, the 3-dimensional metric and
the $2\bar{n}+1$ real, harmonic functions $L_{I}$, $K^{I}$ and $N$.  Given
these ingredients, in order to fully specify the solution we need calculate
$f$, $H$, $\omega$ and $\psi^{I}$ through the following, simplified equations.

\begin{equation}
\begin{array}{lclclcl}
  f^{-3} & =& K_{I}\ K^{I} &\hspace{.3cm},\hspace{.3cm}& L_{I} & =& C_{IJK}K^{J}\ M^{K}\; ,\\
  & & & & & & \\
  H & =& -{\textstyle\frac{1}{2}}L_{I}\ M^{I} \ +\ N & , &
  \hat{\star}\hat{d}\omega & =& \sqrt{3}\left[ L_{I}\hat{d}K^{I} \ -\ K^{I}\hat{d}L_{I}\right] \; ,\\
  & & & & & & \\
  h^{I}(\phi ) & =& f\ K^{I}  & ,&
  \psi^{I} & =& f^3 K^{I}(L_{J}\hat{d}K^{J} -K^{J}\hat{d}L_{J}) +\hat{d}M^{I}\, .   
\end{array}
\end{equation}

Solutions that belong to this family, but depending on a smaller number of
harmonic functions have been given {\em e.g.}~in
Refs.~\cite{Chamseddine:1998yv,Chamseddine:1999xk,Caldarelli:2006ww}.

Apart from being one of the nicest subclasses of solutions, the
$u$-independent one becomes doubleplus interesting when we observe that if we
reduce a solution in the null class over the spacelike direction $\sqrt{2} y =
u-v$, which implies $u$-independence, we end up with a solution in the
timelike class of $N=2$ $d=4$ SUGRA. In fact, comparing the constraints in
this section with the ones in \cite[Sec. (5)]{Hubscher:2006mr}, one finds the
same constraints on the hyperscalars and the 3-dimensional metric.
  
The metric Eq.~(\ref{eq:nullmetric}) can be put in an $y$-adapted system, and
one finds

\begin{equation}
  \begin{array}{rcl}
ds^{2} & = & -k^{2}[dy+A]^{2}
+k^{-1}
\left[
\left(\frac{f^{3}}{1-H}\right)^{1/2}
(dt+{\textstyle\frac{1}{\sqrt{2}}}\omega)^{2}
-\left(\frac{f^{3}}{1-H}\right)^{-1/2}
\gamma_{\underline{r}\underline{s}}dx^{r}dx^{s}
\right]\, ,\\
& & \\
k^{2} & = & (1-H)f\, ,\\
& & \\
A & = & -(1-H)^{-1}(Hdt +{\textstyle\frac{1}{\sqrt{2}}}\omega)\, .\\
\end{array}
\end{equation}

The 4-dimensional solutions can be easily read from these. Apart from the
scalar $k$ and the vector field $A$, which is purely electric if the
5-dimensional solution is static ($\omega=0$), the metric takes the form

\begin{equation}
ds^{2}= \left(\frac{f^{3}}{1-H}\right)^{1/2}
(dt+{\textstyle\frac{1}{\sqrt{2}}}\omega)^{2}
-\left(\frac{f^{3}}{1-H}\right)^{-1/2}
\gamma_{\underline{r}\underline{s}}dx^{r}dx^{s}\, ,  
\end{equation}

\noindent
and belongs to the $N=2,d=4$ timelike class to which all black-hole-type
solutions belong in $d=4$.

This 4-dimensional solution should be compared to the one in
Eq.~(\ref{eq:Naampje}), which is the one one obtains when imposing an extra
isometry on the four dimensional spacelike manifold in the timelike case. the
main difference between them is the electric or magnetic nature of the KK
vector field. In the simplest case this solutions would give a 4-dimensional
electric KK black hole and the other one a 4-dimensional magnetic KK black
hole, related by 4-dimensional electric-magnetic duality, as we discussed in
the introduction. In the more general case, the relation between these
solutions is more complicated and we hope to say more about it in the near
future.


\section{Conclusions}
\label{sec-conclusions}

In this paper we have found new families of supersymmetric solutions with
unfrozen hypermultiplets. These families are very general and the form and
physical properties of each solution depend on the details of the choices of
hypervarieties, harmonic mappings and harmonic functions made. This
opens a new wide range of possibilites that needs to be explored. More work is
need to find out what happens with black hole attractors\footnote{For a
  recent, pedagogical, review, see Ref.~\cite{Larsen:2006xm}.} and black hole
entropy when hyperscalars are unfrozen \cite{kn:BMO}, to find and explain the
generic features of these solutions (are they always singular?), to find out
to which stringy configurations these solutions correspond to etc.

One of the families of solutions describes generically strings with pp-waves
propagating along it and can be dimensionally reduced to supersymmetric
$N=2,d=4$ black holes. This raises new question about how the 4-dimensional
attractor mechanism is implemented in the 5-dimensional setting, taking into
account that these 5-dimensional solutions belong to the null class and the
standard attractor mechanism is proven only for solutions in the timelike
class. The 5-dimensional origin of the 4-dimensional entropy can (and must) be
investigated. 

We hope to report on some of these issues in the near future.


\section*{Acknowledgments}

P.M.~and T.O.~would like to thank S.~Vandoren and A.~Van Proeyen for most
useful comments.  T.O.~would like to thank M.M.~Fern\'andez for her continuous
support.  P.M.~would like to thank C.~Dullemond for his honesty, even though
it took P.M.~10 years to truly appreciate it.  This work has been supported in
part by the Spanish Ministry of Science and Education grant BFM2003-01090, the
Comunidad de Madrid grant HEPHACOS P-ESP-00346, by the {\em Fondo Social
  Europeo} through the I3P programme and by the EU Research Training Network
{\em Constituents, Fundamental Forces and Symmetries of the Universe}
MRTN-CT-2004-005104.

\appendix

\section{Conventions}
\label{sec-d5conventions}

Our conventions can be obtained from those of Ref.~\cite{Bergshoeff:2004kh} by
changing the sign of the metric (to have mostly minus signature), multiplying
all $\gamma^{a}$s by $+i$ and all $\gamma_{a}$s by $-i$ and setting
$\kappa=1/\sqrt{2}$, but we collect here the main features of our conventions
to use them as a reference. In particular,
Section~\ref{sec-realspecialgeometry} contains the relevant Real Special
Geometry identities for $\kappa=1/\sqrt{2}$ (those in Appendix~C of
Ref.~\cite{Bergshoeff:2004kh} are only valid for $\kappa=1$).


\subsection{Gamma matrices and spinors}
\label{app-spinors}
We use mostly minus signature.

The first four of our 5-dimensional gamma matrices are taken to be
identical to 4-dimensional purely imaginary gamma matrices
$\gamma^{0},\gamma^{1},\gamma^{2},\gamma^{3}$ satisfying

\begin{equation}
\{\gamma^{a},\gamma^{b}\}=2\eta^{ab}\, ,  
\end{equation}

\noindent
and the fifth is $\gamma^{4}=-\gamma^{0123}$, so it is purely real,
the above anticommutator is valid for $a=0,\cdots,4$ and, furthermore,
$\gamma^{a_{1}\cdots a_{5}}=+\varepsilon^{a_{1}\cdots a_{5}}$
and, in general

\begin{equation}
\gamma^{a_{1}\cdots a_{n}} =\frac{(-1)^{[n/2]}}{\left(5-n\right)!}
\varepsilon^{a_{1}\cdots a_{n}b_{1}\cdots b_{n-5}}
\gamma_{b_{1}\cdots b_{n-5}}\, .   
\end{equation}

On the other hand, $\gamma^{0}$ is Hermitean and the other gammas are
anti-Hermitean.

To explain our convention for symplectic-Majorana spinors, let us
start by defining the Dirac, complex and charge conjugation matrices
$\mathcal{D}_{\pm},\mathcal{B}_{\pm},\mathcal{C}_{\pm}$. By
definition, they satisfy

\begin{equation}
\mathcal{D}_{\pm}\, \gamma^{a}\, \mathcal{D}_{\pm}^{-1} = 
\pm\gamma^{a\, \dagger}\, , 
\hspace{1cm}
\mathcal{B}_{\pm}\, \gamma^{a}\, \mathcal{B}_{\pm}^{-1} =
\pm\gamma^{a\, *}\, .
\hspace{1cm}
\mathcal{C}_{\pm}\, \gamma^{a}\, \mathcal{C}_{\pm}^{-1} =
\pm\gamma^{a\, T}\, .
\end{equation}

The natural choice for Dirac conjugation matrix is

\begin{equation}
\mathcal{D}=i\gamma^{0}\, ,  
\end{equation}

\noindent
which corresponds to $\mathcal{D}=\mathcal{D}_{+}$. The other
conjugation matrices are related to it by

\begin{equation}
\mathcal{C}_{\pm}=\mathcal{B}^{T}_{\pm}\mathcal{D}\, ,  
\end{equation}

\noindent
but it can be shown that in this case only $\mathcal{C}=\mathcal{C}_{+}$ and
$\mathcal{B}=\mathcal{B}_{+}$ exist and are both antisymmetric. We take them
to be

\begin{equation}
\mathcal{C}=i\gamma^{04}\, ,
\hspace{1cm}  
\mathcal{B}=\gamma^{4}\,\,\,\, \Rightarrow \mathcal{B}^{*}\mathcal{B}=-1\, .
\end{equation}

The Dirac conjugate is defined by

\begin{equation}
        \psi^{\dagger}\mathcal{D}=i\psi^{\dagger}\gamma^{0}\, ,
\end{equation}

\noindent
and the Majorana conjugate by

\begin{equation}
        \psi^{T}\mathcal{C}=i\psi^{T}\gamma^{04}\, .
\end{equation}

\noindent
The Majorana condition (Dirac conjugate = Majorana conjugate) cannot be
consistently imposed because it requires $\mathcal{B}^{*}\mathcal{B}=+1$.
Therefore, we introduce the symplectic-Majorana conjugate in pairs of spinors by
using the corresponding symplectic matrix, {\em e.g.\/}

\begin{equation}
        \psi^{i\, c}\equiv \varepsilon_{ij}\psi^{j\, T}\mathcal{C}\, ,
\end{equation}

\noindent
then the symplectic-Majorana condition is

\begin{equation}
        \psi^{i\, *}=\varepsilon_{ij}\gamma^{4}\psi^{j}\, .
\end{equation}

To impose the symplectic-Majorana condition on hyperinos $\zeta^{A}$ the only
thing we have to do is to replace the matrix $\varepsilon_{ij}$ by
$\mathbb{C}_{AB}$, which is the invariant metric of $Sp(n_h)$.

Our conventions on $SU(2)$ indices are intended to keep manifest the $SU(2)$
covariance. In $SU(2)$, besides the preserved metric, there is the preserved
tensor $\varepsilon_{ij}$. We also introduce $\varepsilon^{ij}$, $\varepsilon_{12} =
\varepsilon^{12} =+1$. Therefore we may construct new covariant objects by using
$\varepsilon_{ij}$ and $\varepsilon^{ij}$, for instance $\psi_{i}\equiv
\varepsilon_{ij}\psi^{j}$ (whence $\psi^{j} = \psi_{i}\varepsilon^{ij}$).
With this notation the symplectic-Majorana condition can be simply stated as

\begin{equation}
        \psi^{i\, *}=\gamma^{4}\psi_{i}\, .
\end{equation}

We use the bar on spinors to denote the (single) Majorana conjugate:

\begin{equation}
        \bar\psi^{i} \equiv {\psi^{i}}^T\mathcal{C}\,,
\end{equation}

\noindent
which transforms under $SU(2)$ in the same representation as $\psi^{i}$ does.
We also lower its $SU(2)$ index: $\bar\psi_{i} \equiv
\varepsilon_{ij}\bar\psi^{j}$.
In terms of single Majorana conjugates the
symplectic Majorana condition reads

\begin{equation}
        \left(\bar\psi^{i}\right)^* = \bar\psi_{i}\gamma^4\,.
\end{equation}

Finally, observe that after imposing the symplectic Majorana condition the
following simple relation between the single Dirac and Majorana conjugates
holds:

\begin{equation}
        {\psi^{i}}^\dagger\mathcal{D} = \bar\psi_{i}\, ,
\end{equation}

\noindent
which is very useful if one prefers to use the Dirac conjugate instead of the
Majorana one.

The bilinears that can be constructed from Killing spinors will in
general be $2\times 2$ matrices that can be written as linear
combinations of the Pauli matrices $\sigma^{\hat r}$ ($\hat{r}=0,\ldots ,3$) where
$\sigma^{0}=\mathbb{I}_{2\times 2}$. Therefore, we are bound to need the Fierz
identities

\begin{equation}
\label{eq:N=2Fierzidentities}
\begin{array}{rcl}
\left(\bar{\lambda}M\varphi\right)\left(\bar{\psi}N\chi\right)
& = & 
\frac{p}{8} \left\{
\left(\bar{\lambda}M\sigma^{\hat{r}} N \chi\right)
\left(\bar{\psi}\sigma^{\hat{r}}\varphi\right)
+
\left(\bar{\lambda}M\gamma^{a}\sigma^{\hat{r}} N \chi\right)
\left(\bar{\psi}\gamma_{a}\sigma^{\hat{r}}\varphi\right)
\right. \\
& & \\
& & 
\left.
-{\textstyle\frac{1}{2}}
\left(\bar{\lambda}M\gamma^{ab}\sigma^{\hat{r}} N \chi\right)
\left(\bar{\psi}\gamma_{ab}\sigma^{\hat{r}}\varphi\right)
\right\}\, ,
\end{array}
\end{equation}

\noindent 
where the $SU(2)$ indices are implicit and $p=(-)1$ for
(anti-)commuting spinors.


\subsection{Spinor bilinears}
\label{sec-bilinears}

With one commuting symplectic-Majorana spinor $\epsilon^{i}$ we can
construct the following independent, $SU(2)$-covariant bilinears:

\begin{enumerate}
\item[$\bar{\epsilon}_{i}\ \epsilon^{j}\;\;\;\;$:] It is easy to see that 

  \begin{equation}
    \begin{array}{rcl}
      \bar{\epsilon}_{i}\epsilon^{j} & = & -\varepsilon^{jk} 
(\bar{\epsilon}_{k}\epsilon^{l}) \varepsilon_{li}\, , \\
& & \\
(\bar{\epsilon}_{i}\epsilon^{j})^{*} & = & 
-\bar{\epsilon}_{j}\epsilon^{i}\, ,\\
\end{array}
\end{equation}

\noindent
The first equation implies that this matrix is proportional to
$\delta_{i}{}^{j}$ and the second equation implies that the constant is
purely imaginary. Thus, we define the $SU(2)$-invariant scalar

\begin{equation}
f \ \equiv\ i \bar{\epsilon}_{i}\epsilon^{i} 
\ =\ 
i \bar{\epsilon}\sigma^{0}\epsilon
\hspace{.5cm},\hspace{.5cm}
\bar{\epsilon}_{i}\epsilon^{j} 
\; =\; 
-\textstyle{i\over 2}\ f\ \delta_{i}{}^{j} \; . 
\end{equation}

\noindent
All the other scalar bilinears $i \bar{\epsilon}\sigma^{r}\epsilon$
($r=1,2,3$) vanish identically.

\item[$\bar{\epsilon}_{i}\gamma^{a}\epsilon^{j}\;$:] This matrix
  satisfies the same properties as $\bar{\epsilon}_{i}\epsilon^{j}$,
  and so we define the vector bilinear

\begin{equation}
V^{a} \ \equiv\ i \bar{\epsilon}_{i}\gamma^{a}\epsilon^{i} 
\ =\ 
i \bar{\epsilon}\gamma^{a}\sigma^{0}\epsilon
\hspace{.5cm},\hspace{.5cm}
\bar{\epsilon}_{i}\gamma^{a}\epsilon^{j} 
\; =\; 
     -\textstyle{i\over 2}\  \delta_{i}{}^{j}\ V^{a} \; .
\end{equation}

\noindent
which is also $SU(2)$-invariant, the other vector bilinears being automatically zero.

\item[$\bar{\epsilon}_{i}\gamma^{ab}\epsilon^{j}$:] In this case

  \begin{equation}
    \begin{array}{rcl}
      \bar{\epsilon}_{i}\gamma^{ab}\epsilon^{j} & = & +\varepsilon^{jk} 
(\bar{\epsilon}_{k}\gamma^{ab}\epsilon^{l}) \varepsilon_{li}\, , \\
& & \\
(\bar{\epsilon}_{i}\gamma^{ab}\epsilon^{j})^{*} & = & 
\bar{\epsilon}_{j}\gamma^{ab}\epsilon^{i}\, ,\\
\end{array}
\end{equation}

\noindent
which means that these 2-form matrices are traceless and Hermitean and
we have three non-vanishing real 2-forms

\begin{equation}
\Phi^{r\, ab} \ \equiv\ 
\sigma^{r}{}_{i}{}^{j}\ \bar{\epsilon}_{j}\gamma^{ab}\epsilon^{i}
\hspace{.5cm},\hspace{.5cm}
\bar{\epsilon}_{i}\gamma^{ab}\epsilon^{j} 
\; =\; 
    \textstyle{1\over 2}{\sigma^{r}}_{i}{}^{j}\ \Phi^{r\, ab} \; .
\end{equation}

\noindent
$r=1,2,3$, which transform as a vector in the adjoint representation of
$SU(2)$, and the fourth $\bar{\epsilon}\gamma^{ab}\sigma^{0}\epsilon=0$.

\end{enumerate}

Using the Fierz identities Eq.~(\ref{eq:N=2Fierzidentities}) for
commuting spinors we get, among other identities, 

\begin{eqnarray}
V^{a}V_{a} & = & f^{2}\, ,\\
& & \nonumber \\
V_{a}V_{b} & = & \eta_{ab}f^{2}
             +{\textstyle\frac{1}{3}}\Phi^{r}{}_{a}{}^{c}\Phi^{r}{}_{cb}\,,\\
& & \nonumber \\
\label{VPhi}
V^{a}\Phi^{r}{}_{ab} & = & 0\, ,\\
& & \nonumber \\ 
\label{VStarPhi}
V^{a}({}^{\star}\Phi^{r}){}_{abc} & = & -f\Phi^{r}{}_{bc}\, ,\\
& & \nonumber \\
\label{quaternions}
\Phi^{r}{}_{a}{}^{c}\Phi^s{}_{cb} & = &  -\delta^{rs}(\eta_{ab}f^{2}
-V_{a}V_{b})-\varepsilon^{rst}f\Phi^t{}_{ab}\, ,\\
& & \nonumber \\ 
\Phi^{r}{}_{[ab}\Phi^s{}_{cd]} & = & -{\textstyle\frac{1}{4}}f\delta^{rs}
\varepsilon_{abcde}V^{e}\, , \\
& & \nonumber   \\
\label{vepsilon}
        V_a\gamma^a\epsilon^{i} &=& f\epsilon^{i}\,,                \\
& & \nonumber   \\
\label{phiepsilon}
\Phi^{r}_{ab}\gamma^{ab}\epsilon^{i} &=& 4if\epsilon^{j}\sigma^{r}{}_{j}{}^{i}\,.
\end{eqnarray}


\subsection{Real Special Geometry}
\label{sec-realspecialgeometry}

The geometry of the $n$ physical scalars $\phi^{x}$
($x=1,\ldots ,n$) of the vector multiplets is fully determined by a
constant real symmetric tensor $C_{IJK}$
($I,J,K=0,1,\ldots,\bar{n}\equiv n+1$).  The scalars appear through
$\bar{n}$ functions $h^{I}(\phi)$ constrained to satisfy

\begin{equation}
  \label{eq:constraint}
C_{IJK}h^{I}h^{J}h^{K}=1\, .  
\end{equation}

\noindent
One defines 

\begin{equation}
h_{I}\equiv C_{IJK} h^{J}h^{K}\, ,\,\,\,\, \Rightarrow h_{I}h^{I}=1\, ,
\end{equation}

\noindent
and a metric $a_{IJ}$ that can be use to raise and lower the
$SO(\bar{n})$ index

\begin{equation}
h_{I}\equiv a_{IJ} h^{J}\, ,
\hspace{1cm}
h^{I}\equiv a^{IJ} h_{J}\, .  
\end{equation}

\noindent
The definition of $h_{I}$ allows us to find 

\begin{equation}
a_{IJ}=-2C_{IJK}h^{K} +3h_{I}h_{J}\, .  
\end{equation}

Next, one defines

\begin{equation}
h^{I}_{x} \equiv -\sqrt{3} h^{I}{}_{,x}\equiv  
-\sqrt{3} \frac{\partial h^{I}}{\partial\phi^{x}}\, ,  
\end{equation}

\noindent
and 

\begin{equation}
h_{Ix}\equiv a_{IJ} h^{J}_{x} =+\sqrt{3}h_{I, x}\, ,
\end{equation}

\noindent
which satisfy

\begin{equation}
h_{I}h^{I}_{x}=0\, ,   
\hspace{1cm}
h^{I}h_{Ix}=0\, ,   
\end{equation}

\noindent
due to Eq.~(\ref{eq:constraint}). The $h^{I}$ enjoy the following
properties of closure and orthogonality

\begin{equation}
\left(
\begin{array}{c} 
h^{I} \\ 
h^{I}_{x} 
\end{array}
\right)
\left(
\begin{array}{cc} 
h_{I} &   h_{I}^{y} \\  
\end{array}
\right)
= 
\left(
\begin{array}{cc}
1 & 0              \\
0 & \delta_{x}^{y} \\
\end{array}
\right)
\, ,
\,\,\,
\left(
\begin{array}{cc} 
h_{I} &  h_{I}^{x} \\
\end{array}
\right)
\left(
\begin{array}{c}
h^{J}     \\
h^{J}_{x} \\
\end{array}
\right)
=
\delta_{I}^{J}\, .           
\end{equation}

Therefore any object with $SO(\bar n)$ index can be decomposed as

\begin{equation}
\label{SOndecomposition}
 A^{I} = \left(h_{J}A^{J}\right)h^{I} + \left(h^{x}_{J}A^{J}\right)h^{I}_{x}\,.
\end{equation}

The metric of the scalars $g_{xy}(\phi)$ is the pullback of $a_{IJ}$:

\begin{equation}
 g_{xy}=a_{IJ}h^{I}_{x}h^{J}_{y}=-2 C_{IJK}h^{I}_{x}h^{J}_{y}h^{K}\, ,
\end{equation}

\noindent
and can be used to raise and lower $x,y$ indices. Other useful
expressions are

\begin{eqnarray}
a_{IJ} & = & h_{I}h_{J}+h^{x}_{I}h_{Jx}\, ,\\
& & \nonumber \\
C_{IJK}h^{K} & = &  h_{I}h_{J}-{\textstyle\frac{1}{2}}h^{x}_{I}h_{Jx}\, ,
\end{eqnarray}

\noindent
and

\begin{eqnarray}
h_{I}h_{J} & = & {\textstyle\frac{1}{3}a_{IJ}} 
+{\textstyle\frac{2}{3}}C_{IJK}h^{K}\, ,\\
& & \nonumber \\
h^{x}_{I}h_{Jx} & = & {\textstyle\frac{2}{3}}a_{IJ} 
-{\textstyle\frac{2}{3}}C_{IJK}h^{K}\, .
\end{eqnarray}

We now introduce the Levi-Civit\`a covariant derivative associated to
the scalar metric $g_{xy}$

\begin{equation}
h_{Ix;y}\equiv h_{Ix,y} -\Gamma_{xy}{}^{z}h_{Iz}\, .  
\end{equation}

\noindent
It can be shown that 

\begin{eqnarray}
h_{Ix;y} & = & 
{\textstyle\frac{1}{\sqrt{3}}}(h_{I}g_{xy} +T_{xyz}h^{z}_{I})\, ,\\  
& & \nonumber \\
h^{I}_{x;y} & = & 
-{\textstyle\frac{1}{\sqrt{3}}}(h^{I}g_{xy} +T_{xyz}h^{Iz})\, ,\\  
& & \nonumber \\
T_{xyz} & = & \sqrt{3} h_{Ix;y}h^{I}_{z} = -\sqrt{3} h_{Ix}h^{I}_{y;z}\, ,\\
& & \nonumber \\
\Gamma_{xy}{}^{z} & = & h^{Iz}h_{Ix,y}
-{\textstyle\frac{1}{\sqrt{3}}}T_{xy}{}^{w} =  
8h_{I}^{z}h^{I}_{x,y}
+{\textstyle\frac{1}{\sqrt{3}}}T_{xy}{}^{w}\, .
\end{eqnarray}


\section{Quaternionic-K\"ahler manifolds}
\label{sec-quaternionic}

In this appendix we review the definition and basics of quaternionic-K\"ahler
manifolds. We refer the reader to Ref.~\cite{Bergshoeff:2002qk} for a more
comprehensive introduction to quaternionic manifolds with original references.

A \textit{quaternionic-K\"ahler manifold} is a real $4n$-dimensional manifold
($n> 1$) such that\footnote{ Clearly, the definitions given below are just too
  weak to be useful when $n=1$, and one defines a 4-dimensional manifold to be
  quaternionic-K\"ahler, iff it is Einstein and selfdual.  For a supergravity
  justification of this definition see {\em e.g.\/} \cite{Bergshoeff:2002qk}.
}

\begin{enumerate}
\item There exists on it a triplet of complex structures $J^{r}{}_{X}{}^{Y}$,
  $r=1,2,3$, $X,Y=1,\ldots 4n$ which satisfy the algebra of imaginary unit
  quaternions,

\begin{equation}
\label{qalgebra}
J^{r} J^{s} \ =\  -\delta^{rs} \ +\ \varepsilon^{rst}\ J^{t}\, ,
\end{equation}

\noindent
which is known as \textit{hypercomplex or quaternionic structure}. A manifold
with this property is an \textit{almost hypercomplex} of almost
\textit{quaternionic manifold}.

\item The hypercomplex structure is integrable, {\em i.e.\/}
  it is covariantly
  constant with respect to the standard Levi-Civit\`a connection and a
  non-trivial $\mathfrak{su}(2)$ connection ({\em i.e.\/}
  with non-vanishing curvature):

\begin{equation}
\partial_{X} J^{r}{}_{Y}{}^{Z}
-\Gamma_{XY}{}^{U} J^{r}{}_{U}{}^{Z} 
+\Gamma_{XU}{}^{Z} J^{r}{}_{Y}{}^{U}
+2\varepsilon^{rst}\omega_{X}{}^{s} J^{t}{}_{Y}{}^{Z} = 0\, ,
\end{equation}

\noindent
where $\omega_{X}{}^{r}$ is the $\mathfrak{su}(2)$ connection.  In this case the
manifold is a \textit{quaternionic manifold}. (If this equation is satisfied
with a trivial $\mathfrak{su}(2)$ connection the manifold is a \textit{hypercomplex
  manifold}.)

\item There is a metric which is invariant under the action of the three
  complex structures

\begin{equation}
g_{XY} = 
J^{(r)}{}_{X}{}^{Z} J^{(r)}{}_{Y}{}^{U} g_{ZU}\, ,
\;\;\;\;\mbox{(no sum over $r$!)}\, .
\end{equation}

\noindent
This property makes it a (quaternionic) K\"ahler manifold.

\end{enumerate}

\noindent

The combination of the complex structures with the metric gives us the three
hyper-K\"ahler 2-forms

\begin{equation}
J^{r}{}_{XY} =g_{XZ}J^{r}{}_{Y}{}^{Z}\, .
\end{equation}

\noindent
They are covariantly closed respect to the $\mathfrak{su}(2)$ connection,

\begin{equation}
dJ^{r} + 2\varepsilon^{rst}\omega^{s}\wedge J^{t} =0\, .
\end{equation}

The holonomy of a quaternionic-K\"ahler manifold is contained in $SU(2)\cdot
Sp(2)$ and the tangent space indices are split accordingly into pairs of
$SU(2)$ and $Sp(n)$ indices $i,j,k=1,2$ and $A,B,C=1,\ldots,2n$ respectively.
The Vielbein is defined to be $f_{iA}{}^{X}$ and is related to the metric
by

\begin{equation}
g_{XY} \; =\;  f_{X}{}^{iA}\ f_{Y}{}^{jB}\ \mathbb{C}_{AB}\ \varepsilon_{ij}\, ,
\end{equation}

\noindent
where 

\begin{equation}
f_{X}{}^{iA}\ f_{iA}{}^{Y} \; =\; \delta_{X}{}^{Y}\, ,\hspace{1cm}
f_{iA}{}^{X}\ f_{X}{}^{jB} \; =\; \delta_{i}{}^{j}\ \delta_{A}{}^{B}\, ,
\end{equation}

\noindent
and $\mathbb{C}_{AB}$ is the $Sp(n)$-invariant metric. The Vielbein also
satisfies the reality condition

\begin{equation}
\left({f_{X}{}^{iA}}\right)^{*} \; =\;  \varepsilon_{ij}\ \mathbb{C}_{AB}\
f_{X}{}^{jB}\, ,
\end{equation}

\noindent
and they are covariantly constant under the combination of the Levi-Civit\`a,
$\mathfrak{su}(2)$- and $\mathfrak{sp}(n)$ connections.  The Vielbein also
gives us the tangent version of the complex structures. The constant matrices
$-i\sigma^{r}$ satisfy the algebra Eq.~(\ref{qalgebra}), and we have

\begin{equation}
J^{r}{}_{X}{}^{Y} \; =\; 
f_{X}{}^{iA}\ \mathtt{J}^{r}{}_{iA}{}^{jB}\ f_{jB}{}^{Y}\, , 
\hspace{1cm}
\mathtt{J}^{r}{}_{iA}{}^{jB} \; \equiv\; 
-i\sigma^{r}{}_{i}{}^{j}\ \delta_{A}{}^{B}\, .
\end{equation}

The spin connection can be split into its $\mathfrak{su}(2)$ and $\mathfrak{sp}(n)$
components as follows:

\begin{equation}
\label{eq:spinsplit}
\omega_{X\, iA}{}^{jB} \; =\; 
{\textstyle\frac{i}{2}}\ \omega_{X}{}^{r}\ \mathtt{J}^{r}{}_{iA}{}^{jB}
\; +\; \omega_{X\, A}{}^{B}\ \delta_{i}{}^{j}\, .
\end{equation}

Some useful identities are 

\begin{eqnarray}
\label{eq:una}
R_{XY}{}^{r} & = & {\textstyle\frac{1}{4}}\nu\ J^{r}{}_{XY}\, ,\\
& & \nonumber \\
2f_{[X}{}^{iA}f_{Y]jA} & = & 
iJ^{r}{}_{XY}\ \sigma^{r}{}_{j}{}^{i}\, ,\\
& & \nonumber \\
\label{eq:otra}
2f_{(X}{}^{iA}f_{Y)jA} & = & g_{XY}\ \delta_{j}{}^{i}\, .
\end{eqnarray}

\noindent
The constant $\nu$ is given in terms of the dimensionality of the manifold
$4n$ and its Ricci scalar $R$ by 

\begin{equation}
\nu = \frac{R}{4n(n+2)}\, .  
\end{equation}


\section{The $d=5$ conformastationary metric}
\label{app-metric}

In the timelike case we find the conformastationary metric
Eq.~(\ref{conforma-stationary} ) which we rewrite here for convenience:

\begin{equation}
ds^{2} = f^{2}\left(dt+\omega\right)^{2}
-f^{-1}h_{\underline{m}\underline{n}} dx^{m}dx^{n}\, ,
\hspace{1cm}
\omega=\omega_{\underline{m}}dx^{m}\, ,
\hspace{1cm}
m,n=1,\cdots,4\, .
\end{equation}

\noindent
We choose the Vielbein basis

\begin{equation}
(e^{a}{}_{\mu}) = 
\left(
  \begin{array}{cc}
f & f \omega_{\underline{m}} \\
& \\
0 & f^{-1/2} V^{n}{}_{\underline{m}} \\
  \end{array}
\right)\, ,
\hspace{1cm}
(e^{\mu}{}_{a}) = 
\left(
  \begin{array}{cc}
f^{-1} & -f^{1/2} \omega_{m} \\
& \\
0 & f^{1/2} V^{\underline{n}}{}_{m} \\
  \end{array}
\right)\, ,
\end{equation}

\noindent
where 

\begin{equation}
h_{\underline{m}\underline{n}}
=V_{\underline{m}}{}^{p}V_{\underline{n}}{}^{q}\delta_{pq}\, ,
\hspace{1cm}
V_{m}{}^{\underline{p}}\, V_{n}{}^{\underline{q}}\, 
h_{\underline{p}\underline{q}}
=\delta_{mn}\, ,
\hspace{1cm}
\omega_{m}= V_{m}{}^{\underline{n}}\omega_{\underline{n}}\, .   
\end{equation}

The non-vanishing components of the spin connection in this basis are

\begin{equation}
\label{eq:conformaspincon}
\omega_{00m}= -2\partial_{m}f^{1/2}\, ,
\hspace{.5cm}
\omega_{0mn}=\omega_{m0n}=
{\textstyle\frac{1}{2}}f^{2}\left(d\omega\right)_{mn}\, ,
\hspace{.5cm}
\omega_{mnp}=-f^{1/2}\xi_{mnp}
-2\delta_{m[n}\partial_{p]}f^{1/2}\, ,
\end{equation}

\noindent
where, from now on, all the objects in the r.h.s.~of these equations refer to
the 4-dimensional metric $h_{\underline{m}\underline{n}}$ and, in particular

\begin{equation}
\left(d\omega\right)_{mn} =  
V_{m}{}^{\underline{p}}  V_{n}{}^{\underline{q}}
\left(d\omega\right)_{\underline{p}\underline{q}}=
2V_{m}{}^{\underline{p}}  V_{n}{}^{\underline{q}}
\partial_{[\underline{p}}\omega_{\underline{q}]}\, .
\end{equation}

Thee non-vanishing components of the Ricci tensor are

\begin{equation}
  \begin{array}{rcl}
R_{00} & = & -\nabla^{2}f +f^{-1}(\partial f)^{2}
-\frac{1}{4}f^{4}(d\omega)^{2}\, ,\\
& & \\
R_{0m} & = & -\frac{1}{2}f^{-1/2}\nabla_{n}[f^{3}(d\omega)_{nm}]\, ,\\
& & \\
R_{mn} & = & fR_{mn} -\frac{1}{2}(d\omega)_{mp}(d\omega)_{np}
+\frac{3}{2}f^{-1}\partial_{m}f\partial_{n}f 
-\frac{1}{2}\delta_{mn}[\nabla^{2}f -f^{-1}(\partial f)^{2}]\, ,\\
  \end{array}
\end{equation}

\noindent
and the Ricci scalar is given by

\begin{equation}
R= -fR +{\textstyle\frac{1}{4}}(d\omega)^{2} +\nabla^{2}f 
-{\textstyle\frac{5}{2}}f^{-1}(\partial f)^{2}\, .
\end{equation}

We define, following Ref.~\cite{Gauntlett:2002nw} we define the decomposition 

\begin{equation}
fd\omega = G^{+}+ G^{-}\, ,
\end{equation}

\noindent
so

\begin{equation}
de^{0}=f^{-1}df\wedge e^{0}+G^{+}+G^{-}\, .
\end{equation}

\noindent
Further, since in this basis $\hat{V}=fe^{0}$, we have

\begin{equation}
  \begin{array}{rcl}
d\hat{V} & = & 2df\wedge e^{0} +f(G^{+}+G^{-})\, ,\\
& & \\
\star d\hat{V} & = & 2\star (df\wedge e^{0})+(G^{+}-G^{-})\wedge \hat{V}\, .\\
  \end{array}
\end{equation}


\section{The null-case metric}
\label{app-nullmetric}

\begin{equation}
ds^{2}= 2fdu(dv+Hdu+\omega)
-f^{-2}\gamma_{\underline{r}\underline{s}}dx^{r}dx^{s}\, ,
\hspace{1cm}
r,s=1,2,3\, .  
\end{equation}

\noindent
Orthonormal 1-form and vector basis for this metric are given by

\begin{equation}
  \begin{array}{rclrcl}
e^{+} & = & fdu\, ,& e_{+} & = & 
f^{-1}(\partial_{\underline{u}} -H \partial_{\underline{v}})\, ,\\
& & & & & \\ 
e^{-} & = & dv+Hdu+\omega\, ,\hspace{1.5cm}& e_{-} & = & \partial_{\underline{v}}\, ,\\
& & & & & \\ 
e^{r} & = & f^{-1}v^{r}\, , & e_{r} & = & f (v_{r} -\omega_{r}
\partial_{\underline{v}})\, ,\\
  \end{array}
\end{equation}

\noindent
where $v^{r}=v^{r}{}_{\underline{s}}dx^{s}$ and
$v_{r}=v_{r}{}^{\underline{s}}\partial_{\underline{s}}$ are orthonormal basis
1-forms and vectors for the 3-dimensional spatial positive-definite metric
$\gamma_{\underline{r}\underline{s}}$

\begin{equation}
\delta_{rs}v^{r}{}_{\underline{t}} v^{s}{}_{\underline{q}}
=\gamma_{\underline{t}\underline{q}}\, ,
\hspace{1cm}
v_{t}{}^{\underline{r}}
v_{q}{}^{\underline{s}}
\gamma_{\underline{r}\underline{s}} 
=\delta_{tq}\, .
\end{equation}

The non-vanishing components of the spin connection are 

\begin{equation}
  \begin{array}{rclrcl}
\omega_{+r+} & = & \partial_{r}H 
-\partial_{\underline{u}}\omega_{\underline{s}}
v_{r}{}^{\underline{s}}\, ,& 
\omega_{rs+} & = & -\frac{1}{2}f^{2} F_{rs}  
-f^{-2}\partial_{\underline{u}}f \delta_{rs} 
-f^{-1}v_{(r|}{}^{\underline{t}}
\partial_{\underline{u}} v_{|s)\underline{t}}\, ,\\
& & & & & \\
\omega_{+r-} & = & \frac{1}{2}\partial_{r}f = 
\omega_{-r+} = -\omega_{r+-}\, ,&
\omega_{+rs} & = & \frac{1}{2}f^{2} F_{rs}
-f^{-1}v_{[r|}{}^{\underline{t}}
\partial_{\underline{u}} v_{|s]\underline{t}}\, ,\\ 
& & & & & \\
\omega_{rst} & = & f \varpi_{rst} -2\delta_{r[s}\partial_{t]}f\, ,& & & \\ 
\end{array}
\end{equation}

\noindent
where all the quantities in the r.h.s.~of all these equations refer to the
3-dimensional metric and Dreibein and 

\begin{equation}
F_{rs} = v_{r}{}^{\underline{t}} v_{s}{}^{\underline{p}}
F_{\underline{t}\underline{p}}\, ,
\hspace{1cm}
F_{\underline{r}\underline{s}}\equiv 
2\partial_{[\underline{r}}\omega_{\underline{s}]}\, .   
\end{equation}

The non-vanishing components of the Ricci tensor are

\begin{equation}
  \begin{array}{rcl}
R_{++} 
& = & 
-f\nabla^{2} H -{\textstyle\frac{1}{4}}f^{4}F^{2} 
+f\nabla^{\underline{r}}\dot{\omega}_{\underline{r}} 
+ 3\dot{\omega}_{\underline{r}}\partial^{\underline{r}}f 
+{\textstyle\frac{1}{2}}f^{-2}\gamma^{\underline{rs}}
\ddot{\gamma}_{\underline{rs}} 
+{\textstyle\frac{1}{4}}f^{-2}\dot{\gamma}^{\underline{rs}}
\dot{\gamma}_{\underline{rs}} 
\\
& & \\
& & -{\textstyle\frac{3}{2}} f^{-3}\dot{f} \gamma^{\underline{rs}}
\dot{\gamma}_{\underline{rs}}
-3f^{-2}\left[\partial_{\underline{u}}^{2}\log{f} 
-2\left(\partial_{\underline{u}}\log{f}\right)^{2}\right]\, ,
\\
& & \\
R_{+-} 
& = & 
-{\textstyle\frac{1}{2}}f^{2}\nabla^{2}\log{f}\, ,
\\
& & \\
R_{+r} 
& = & 
-{\textstyle\frac{1}{2}}\nabla_{s} \left(f^{3}F_{sr}\right) 
-{\textstyle\frac{1}{2}}v_{r}{}^{\underline{r}}\gamma^{\underline{st}}
\nabla_{\underline{s}}\dot{\gamma}_{\underline{rt}} 
+{\textstyle\frac{1}{2}}v_{r}{}^{\underline{r}}\partial_{\underline{u}}
\left(\gamma^{\underline{st}}\partial_{\underline{r}}
\gamma_{\underline{st}}\right)
+{\textstyle\frac{3}{2}}v_{r}{}^{\underline r} \dot{\gamma}_{\underline{rt}}
\partial^{\underline{t}}\log{f} \\
& & \\
& &
-{\textstyle\frac{3}{2}}\partial_{r}\partial_{\underline u}\log{f}
-{\textstyle\frac{3}{4}}\gamma^{\underline{st}} \dot{\gamma}_{\underline{st}}
\partial_{r}\log{f}
+{\textstyle\frac{3}{2}}\partial_{\underline{u}}\log{f}\partial_{r}\log{f}\, ,
\\
& & \\
R_{rs} 
& = & 
f^{2}R_{rs}(\gamma) -\delta_{rs}f^{2}\nabla^{2}\log{f} 
+{\textstyle\frac{3}{2}}\partial_{r} f \partial_{s} f\, ,\\  
      \end{array}
\end{equation}

\noindent
and  the Ricci scalar is

\begin{equation}
R = -f^{2} R(\gamma) + 2f^{2}\nabla^{2}\log{f} 
-{\textstyle\frac{3}{2}}\left(\partial f\right)^2\, .
\end{equation}


\end{document}